\newcommand{\ergs}{erg\,s$^{-1}$}
\newcommand{\kms}{km\,s$^{-1}$}
\newcommand{\Zsol}{$Z_{\odot}$}
\newcommand{\Msol}{$M_{\odot}$ }
\newcommand{\Macc}{$\dot{M}_{acc}$}

\newcommand{\gcas}{\object{$\gamma$ Cas}}
\newcommand{\hd}{\object{HD 110432}}
\newcommand{\hda}{\object{HD 161103}}
\newcommand{\hdb}{\object{HD 119682}}
\newcommand{\sao}{\object{SAO 49725}}
\newcommand{\usno}{\object{USNO 0750--13549725}}
\newcommand{\ssbe}{\object{SS 397}}

\newcommand{\FeHe}{\ion{Fe}{xxv}}
\newcommand{\FeH}{\ion{Fe}{xxvi}}
\newcommand{\FeK}{Fe\,K$\alpha$}

\newcommand{\obsA}{\textsc{obs.} {\small 1}}
\newcommand{\obsB}{\textsc{obs.} {\small 2}}
\newcommand{\obsC}{\textsc{obs.} {\small 3}}

\newcommand{\XMM}{XMM-\textit{Newton}}

\documentclass{aa} \usepackage{natbib,graphicx}  \usepackage{natbib,graphicx} 

\begin{document}


\title{On the X-ray and optical properties of the Be star
\hd:\\ a very hard-thermal X-ray emitter\thanks{Based on the public data archive of the XMM-{\it 
Newton}, an ESA science mission with instruments and contributions directly funded by ESA Member
States and NASA.
}}

\author{R. Lopes de Oliveira\inst{1,2} \and C. Motch\inst{2} \and M.A. Smith\inst{3} \and I. Negueruela\inst{4} \and J.M. Torrej\'{o}n\inst{4}}

\offprints{R. Lopes de Oliveira,\\
\email{rlopes@astro.iag.usp.br}}

\institute{ 
Instituto de Astronomia,
Geof\'{\i}sica e Ci\^encias Atmosf\'ericas, Universidade de S\~ao Paulo, R. do
Mat\~ao 1226, 05508-090 S\~ao Paulo, 
Brazil \and Observatoire Astronomique, UMR 7550 CNRS, Universit\'e Louis Pasteur, 11 rue de l'Universit\'e, F-67000 Strasbourg, France \and Catholic University of America, 
3700 San Martin Drive, Baltimore, MD 21218 \and Departamento de F\'{\i}sica, Ingenier\'{\i}a de Sistemas y Teor\'{\i}a de la Se\~nal, Escuela 
Polit\'ecnica Superior, Universidad de Alicante, Ap. 99, 03080 Alicante, Spain}

\date{Received 14 February 2007 / Accepted 23 August 2007}

\authorrunning{Lopes de Oliveira et al.}  \titlerunning{On the X-ray and optical properties of \hd}

\abstract{
\hd\  is the first proposed, and best studied, member of a growing group of Be stars with X-ray properties similar to \gcas. 
These stars exhibit hard-thermal X-rays that are variable on all
measurable timescales. 
 This emission contrasts with the soft emission of ``normal" massive stars and
with the nonthermal  emission of all well known Be/X-ray binaries -- so far, all Be + neutron star systems.
In this work we present X-ray spectral and timing properties of \hd\ from three \XMM\ observations in addition to new optical spectroscopic observations.
Like \gcas, the X-rays of \hd\ appear to have a thermal 
origin, as supported by strongly ionized \FeHe\ and \FeH\ lines detected 
in emission. 
A fluorescent iron feature at 6.4 keV is present in all observations, 
while the \FeH\ Ly$\beta$ line is present in two 
of them.
Its X-ray spectrum, complex and time variable, is well described in
each observation by 
three thermal plasmas with temperatures ranging between 0.2--0.7, 3--6, 
and  16--37 keV. Thus, \hd\ has the hottest thermal plasma of any known Be star. 
A sub-solar iron abundance ($\sim$ 0.3--0.5$\times Z_{Fe,\odot}$) is derived 
for the hottest plasma, while lines of less excited ions at longer wavelengths are consistent 
with solar abundances.
The star has a moderate 0.2--12 keV luminosity of $\sim$ 
5$\times$10$^{32}$\,erg\,s$^{-1}$. 
The intensity of the X-ray emission is strongly variable. Recurrent 
flare-like events on time scales as short as $\sim$ 10 seconds are superimposed 
over a basal flux which varies on timescales of $\sim$ 5--10$\times$10$^{3}$ seconds, followed by similarly rapid 
hardness variabilities.
There is no evidence for coherent oscillations, and an upper 
limit of $\sim$ 2.5\% is derived on the pulsed fraction for short pulsations from 0.005 to 2.5\,Hz.
In the optical region the strong and quasi-symmetrical profile of the 
H$\alpha$ line (EW $\sim$ -60 \AA) as well as the detection of several metallic 
lines in emission strongly suggest a dense and/or large circumstellar disk. 
Also, the double-peaked profiles of metallic lines confirm the nearly edge-on 
projection of that disk noted recently by Smith \& Balona. 
\hd\ has several properties reminiscent of the cataclysmic variables such
as a very hot X-ray temperature and some of its detailed spectral features. This suggests
that it might be a Be star harbouring an accreting white dwarf.
On the other hand, recent evidence of magnetic activity reported in the literature of \hd\ suggests an interaction between the surface of the Be star and its disk can produce the X-rays.

 }

\maketitle

\section{Introduction}\label{introduction}

Although \gcas\ (B0.5\,Ve) has long stood out as having unique X-ray properties
among massive stars, accumulating evidence suggests that it is only the first
of a growing subclass of X-ray active Be stars.  
\hd\ \citep[B0.5-1\,III-IVe;][]{Codina84,Dachs86,SB06} was the first proposed  \gcas\ analog on the basis of its mean
2--10 keV fluxes and energy distribution, and optical and X-ray
variability properties \citep{RSH02,SB06}.  Other similar objects are
\hda, \sao, \usno, \ssbe\ \citep{Motch97,Motch06a,Lopes06}, and \hdb\
\citep{R06,SH07}.
Altogether, these objects point to a new class of X-ray emitters \citep{Motch06a,Lopes06} composed of Be stars having an unusual hard-thermal X-ray emission ($k$T $\ga$ 7 keV) of moderate 
luminosity (a few 10$^{32}$ \ergs, at 0.2--12 keV), which displays marked
variability on long and short time scales.
However, in contrast with the behaviour of many
Be/X-ray systems, no outburst has yet been observed from any \gcas\ system.
Curiously, all known members are B0.5e-B1e stars.

The nature of the X-ray emission of \gcas\ and its analogs is 
currently a matter of controversy. Two exciting interpretations 
have been proposed in the recent literature: [1] single-Be stars with 
unusually strong magnetic activity, and tentatively associated to 
progenitors of magnetars, [2] binary systems with an accreting 
degenerate companion -- most likely Be + white dwarf (WD) systems, predicted by the evolutionary models of massive binary systems but still not identified, according to the thermal nature of their X-rays \citep[see discussion in][]{Lopes06}.

In order to understand the nature of these objects and eventually unveil the
mechanism leading to their unusual X-ray emission, compared to that of ``normal"
B/Be stars and known Be/X-ray systems, we have started an extensive X-ray and
optical program intended to search for, and study new \gcas\ analogs.
Here we report on the X-ray properties of \hd\ based on a campaign of three
EPIC/XMM-{\it Newton} observations, along with optical observations. \hd's properties are
discussed in terms of single-star and accreting binary models.

\begin{table*} \caption{Journal of \XMM\ observations. In all cases the {\it thick} filter was applied to reject 
optical light, and the {\it pn} and MOS1-2 cameras were operated in the {\it extended full window} and {\it full 
window} modes, respectively.}              
\label{tab:obslog}        \centering           \begin{tabular}{c c c c c c c c c c c c c}           

\hline                    
\hline\\[-2.2ex]

Obs. ID		& EPIC	& Date & Start time 		&   Duration    & Source counts$^{\mathrm{a}}$  	   & 
GTI$^{\mathrm{b}}$  \\
		&	Cameras	& &			&   (s)        &  (cts s$^{-1}$) / (\% of the events) & (\% of the obs.)\\
\hline\\[-2.2ex]
0109480101	& {\it pn}	& 3 July 2002 & T15:51:56 	& 49398 	& 2.726 / 98.2\%	  & 98.4\% 
\\
(\obsA)	& MOS1	 	&  & T15:00:31       & 52778	     & 0.740 / 97.3\%       &  100\%\\
		& MOS2	 	& &   T15:00:31       & 52778	     & 0.994 / 97.2\%       &  100\%\\
\hline\\[-2.2ex]
0109480201	& {\it pn}	& 26 August 2002 & T21:55:14     & 44823	      & 2.269 / 95.0\%	&  52.6\%\\
(\obsB)	& MOS1   	&  & T21:03:49     & 47827		& 0.911 / 95.8\%	  &  93.2\%\\
		& MOS2   	& &  T21:03:50     & 47837		& 0.932 / 96.0\%	  &  92.4\%\\
\hline\\[-2.2ex]
0109480401	& {\it pn}	& 21 January 2003 & T01:07:38    & 44391	      & 1.913/ 97.2\% 	&  97.8\%\\
(\obsC)	& MOS1   	& &  T00:16:20    & 47764	       & 0.663 / 96.5\%		 &  99.2\%\\
		& MOS2   	& &  T00:16:12    & 47772	       & 0.644 / 96.9\%	 &  99.8\%\\ 

\hline                                  \end{tabular} 
\begin{list}{}{}
\item[$^{\mathrm{a}}$] data at 0.6--12 keV collected during all observation time, corrected on axis; $^{\mathrm{b}}$ good time interval, after excluding times of enhanced soft-proton background: $<$\,1 (and $<$\,0.4) counts per second in the single-events {\it pn} 
(and MOS) light curve from whole camera, at E $>$ 10 keV.
\end{list}
\end{table*}

\section{Previous observations of the main properties of \hd}

\subsection{From the near-infrared to UV}

\hd\ is a bright B1IVe star ($V$\,=\,5.2, $B$\,=\,5.5), located behind the
Southern Coalsack, probable member of the open cluster \object{NGC\,4609} 
\citep[$\sim$ 60\,Myr;][]{Codina84,Feinstein71,Kilkenny85,Kharchenko05,SB06}. 
The UV-to near IR spectrum and the Balmer jump of \hd\ imply a T$_{\rm eff}$ 
of 25000\,K and 22510\,K, and $log\,g$ of 3.5 and 3.9, respectively 
\citep{Codina84,Zorec05}. \citet{Zorec05} estimated a mass of 9.6\,\Msol 
for the star. \hd\ is not known to be in a binary system. 
Rotational velocity $V_{rot}sin$ $i$ measures derived 
using different optical and UV lines vary from 300 to 400\,\kms\ 
\citep{Slettebak82,Codina84,Ballereau95,Fremat05,SB06}. 
For example, the \ion{He}{ii} $\lambda$1640 profile is similar in 
equivalent width and broadening to the feature in the \gcas\ spectrum.
High resolution spectra obtained of \hd\ 
by \citet{SB06} in 2005 January--February
show that the \ion{He}{i} $\lambda$4471, $\lambda$5876, and $\lambda$6678
lines include strong wings extending to at least $\pm$ 1000 km\,s$^{-1}$.
The cause of these wings is unknown. 
Although single previous observation of $\lambda$4471 published by 
\citet{Ballereau95} did not cover this full velocity range, their spectrum of this line in this object exhibited
emission bumps to the blue and red of the line center that have not
been reported in other spectra of the star. 
The \ion{He}{i} and strong metallic lines in the yellow-red region shows
symmetrical emission peaks spaced $\pm$ 100--115 km\,s$^{-1}$ from line
center. 
Similar features are observed in the near-infrared region, in the \ion{He}{i} $\lambda$10830 and H$_{\rm Pa \gamma}$ $\lambda$10938 lines \citep[peaks at $\pm$ 80 km\,s$^{-1}$;][]{Groh07}. Metallic emission features were 
also reported in photographic spectra of \gcas\ \citep{Bohlin70}. 
The kinematic separation and strengths of the peaks in \hd\ indicate that
the disk is viewed nearly edge on and that its mass and geometric extent
are on the ``high end" of what is typical for classical Be stars.
The presence of a strong H$\alpha$ line was reported by \citet{Dachs86}. The EW was $\sim$ -49\AA\ in 1982 and -52.3\AA\ in 1983.

  Both optical spectroscopic and photometric variability is also prevalent 
in \hd. \citet{SB06} found that the \ion{He}{i} $\lambda$5876 and $\lambda$6678
lines often exhibit ``migrating sub-features" ({\it msf}), which are narrow absorptions moving blue-to-red across the line profile.
These appear in line profiles at irregular intervals and 
proceed across the profile with an acceleration 
near 100 km\,s$^{-1}$\,hr$^{-1}$. Such features have been observed by
several authors in the \gcas\ spectrum \citep{Yang88,Smith95,SR99}. 

  A photometric period of either 1.77 or 1.42 days has been claimed for \hd\
by \citet{Barrera91} on the basis of 34 observations, but
we cannot confirm this on the basis of photometric monitoring in 2002,
as reported by \citet{SB06}. However, both the Cousins $B$ and $V$-band 
light curves of this paper exhibited a clear $\sim$ 3--4\% 
sinusoidal modulation with a timescale of 130 days in 2002.

\subsection{X-ray properties}

BeppoSAX follow-up observations \citep{TO01} of some hard X-ray sources discovered in the HEAO-1 all-sky survey \citep{Tuohy88} revealed the unusual X-ray emission of \hd. The X-ray energy distribution was well described by a hot-thermal plasma model with $k$T $\sim$ 11 keV or by a {\it power law} model with photon index $\Gamma$\,$\sim$\,1.63 and a cut-off energy E$_{c}$ $\sim$ 19.9 keV. This last model required additional gaussian lines at 6.76 keV and 8.4 keV.  
The first line, with an EW of $\sim$ 600--700 eV, was interpreted as an unresolved blend of the \FeHe 
(6.7 keV) and \FeH\ (6.97 keV) lines. As this feature is naturally predicted by the hot plasma model, a thermal 
interpretation of the X-ray emission was logically preferred.
The BeppoSAX data did not show evidence for the presence of a fluorescent emission iron line at 6.4 keV.
In the 2--10 keV band, the X-ray flux corrected for absorption ($\sim$ 3.2$\times$10$^{-11}$\,erg\,cm$^{-2}$\,s$^{-1}$) yields a luminosity of 3.4$\times$10$^{32}$\,\ergs\ if the distance to the system is 300\,pc \citep[Hipparcos; ][]{Perryman97}.
The X-ray spectrum of \hd\ was affected by a hydrogen column of $\sim$ 1.1--1.4$\times$10$^{22}$\,cm$^{-2}$, more than the $\sim$ 2$\times$10$^{21}$\,cm$^{-2}$ value due to the Galactic reddening toward the star as derived from the colour excess 
\citep[E($B$-$V$)\,=\,0.4;][]{Rachford01}. A single ``oscillation'' with a 
timescale of 14\,ks was also suspected by \citet{TO01}. 
These authors concluded that if this variation could be confirmed as periodic,
it would provide evidence for \hd\ being a high mass X-ray binary with
an accreting WD companion.
In their picture the X-ray variations
could be caused by rotational modulation of a hot spot on this companion.
\citet{SB06} noted that the fluctuations similar to this single modulation
also occur in the light curve of \gcas\, but these do not repeat regularly.

\section{Observations}

\subsection{Optical observations}

We observed \hd\ using the ESO Multi-Mode Instrument (EMMI) on the
3.5-m New Technology Telescope (NTT) at La Silla, Chile, on two
occasions. On 5 June 2003, an H$\alpha$ spectrum was taken with the red
arm in intermediate-resolution mode (REMD) and grating \#6. The red arm is
equipped with a mosaic of two thin, back-illuminated $2048\times4096$
MIT/LL CCDs and this configuration results
in a nominal dispersion of 0.4\AA/pixel over the range
$\lambda\lambda$6440--7150\AA. The resolution, measured from the widths of arc lines, is
$\sim 1.2$\AA. The blue spectrum was taken with the blue arm in
intermediate-resolution mode (BLMD) and grating \#12. The blue arm is
equipped with a Textronik TK1034 thinned, back-illuminated
$1024\times1024$ CCD and this configuration results in a nominal
dispersion of 0.9\AA/pixel over the range $\lambda\lambda$3820--4750\AA.
The resolution, measured on arc lines, is $\sim 2.6$\AA.

 On 10 May 2004, we observed the H$\alpha$ region with the same
configuration, but a slightly different grating angle, resulting in
coverage of the $\lambda\lambda$6170--6880\AA\ range. The blue spectrum
was taken with the red arm in exactly the same configuration, but using
grating \#7 instead. (The red arm has lower efficiency in the blue, but
grating \#7 allows a much better compromise between resolution and range
than any blue grating). The spectrum covers the range
$\lambda\lambda$3800--5200\AA, with a nominal dispersion of 0.85
\AA/pixel and a measured resolution of $\sim 2.5$\AA.

Image pre-processing was carried out with {\em MIDAS}
software, while data reduction was achieved with the {\em Starlink}
packages {\sc ccdpack} \citep{Draper00} and {\sc figaro}
\citep{Shortridge97}.

\subsection{X-ray observations}

\hd\ was serendipitously observed in a campaign of three \XMM\ exposures of the Wolf-Rayet star \object{WR 47} (see Table~\ref{tab:obslog} for details). Its extreme off-axis position (in the range of 13\arcmin) does not allow it to be in the field of view of the high-resolution RGS cameras.
Only the medium-resolution (E/$\Delta$E $\sim$ 20--50) EPIC cameras (MOS1, MOS2, and $pn$) were able to detect \hd\ in its fields, which cover an area of $\sim$ 700\,arcmin$^{2}$ in the used observational modes ({\it extended full window} for $pn$, and {\it full 
window} for MOS1 and MOS2).
The observations were taken on 3 July 2002 (hereafter \obsA), 26 August 2002 (\obsB), and 21 January 2003 (\obsC).

\begin{figure*} \centering{ 
\includegraphics[bb=1.7cm 13.5cm 20cm 23.8cm,clip=true,width=18cm]{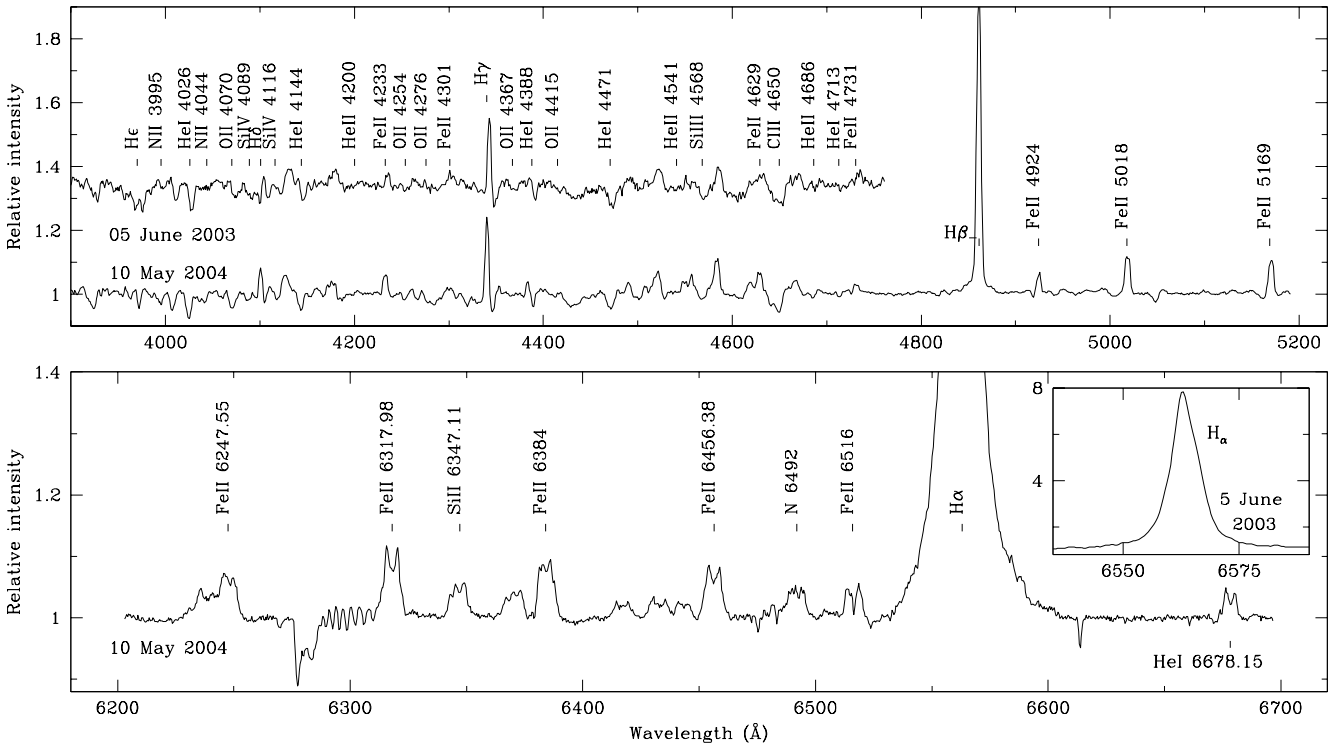}
\caption{Blue and red optical spectra of \hd. The inset shows the profile of the H$\alpha$ line.} 
\label{fig:optspec}  } \end{figure*}

We use the {\it Science Analysis Software} (SAS) v6.5 and the most recent calibration files for the data processing. The spectral analysis was performed with \textsc{Xspec}\,v11.3.1.

In the following only single and double event patterns ($\leq$\,4), and single to quadruple event patterns ($\leq$\,12) were used for {\it pn} and MOS cameras, respectively. 
The only exception is in evaluating the parameters of the iron lines (Table \ref{tbl:EW_FeKa} and Fig. \ref{fig:fekalpha}) in which only single-{\it pn} events collected during the integrated observation were used in order to obtain the best possible energy resolution ($\sim$ 150 eV at 6.4 keV)\footnote{XMM-{\it Newton} User's Handbook, issue 2.4.}. According to the {\it epatplot}/SAS task, the data are not affected by photon pile up.

Only a few time intervals were affected by slightly enhanced background of solar particles (Table~\ref{tab:obslog}).
We exclude photons collected during these intervals when investigating the spectral energy distribution, and we use the entire observation for the timing analysis and investigation of parameters of the \FeK\ complex.

There is a residual shoulder around 0.3 keV in the   spectra which could be due to bad cancellation of a strong absorption feature present in the thick filter.  For that reason, we preferred to discard the low energy part of the spectrum below 0.6 keV.  The background fluxes were extracted in large
regions located in the same CCD as the source. For spectral studies, the energy channels of each camera were grouped in bins containing at least 150 events, and the \textsc{phabs} model was applied to account for the photoelectric absorption.

Throughout this paper, we use data collected by all EPIC cameras (MOS1, MOS2, and {\it pn}) in timing analysis, but owing to possible
uncertainties in the cross-calibration between the three cameras at large off-axis angle we opted to use only the $pn$ data for our spectroscopic analysis.
We checked that the lower S/N MOS data yielded spectral
parameters consistent with those derived from the EPIC $pn$.

\section{Optical properties}

There are numerous metallic emission lines in the blue and red spectra of \hd\ (Fig. \ref{fig:optspec}), chiefly arising from \ion{Fe}{ii} transitions, in addition to strong H-Balmer lines.
As for a number of Be stars \citep{Slettebak92}, the detection of these lines in \hd\ suggests the presence of a dense and/or large circumstellar disk, and according to the double-peak feature of the metallic lines, in a nearly edge-on orientation.
We derive an equivalent width (EW) of -60$\pm$1\AA\ for the H$\alpha$ line.

Interestingly, the double-lobe emissions of the metallic lines seem more equal in Fig. \ref{fig:optspec} than those obtained by \citet{SB06}, in which on average the blue component was slightly stronger. We speculate that some kind of
instability in the \hd's disk is excited --
most likely an one-arm disk instability \citep[e.g.][]{Okazaki91}.

The presence of the emission lines renders difficult the identification of weak absorption features, and therefore the determination of the spectral type and luminosity class.
The spectral type can be estimated from the presence of a moderately
strong \ion{Si}{iv}~4089\AA\ line and a weak \ion{He}{ii}~4686\AA\
line to be close to B0.5-1, while the
relative strength of the complex around \ion{C}{iii}~4650\AA\
suggests that it is likely to be a moderately luminous star. 
Thus, we conclude that the luminosity class of \hd\ is compatible with III or IV.

\section{The spectral energy distribution in X-rays}
\label{sect:spectra}

\subsection{The epochal spectra}
\label{sect:AvSpct}

The \XMM\ observations of \hd\ were investigated separately for each epoch (as Table \ref{tab:obslog}), because its X-rays are clearly dependent on epoch. For example, we find that the local absorption(s) varies between different observations. The star exhibits a hard X-ray spectrum, in which an \FeK\ complex in emission -- composed of a fluorescent feature at 6.4 keV, \FeHe\ (He-like) at 6.7 keV, and \FeH\ (H-like) at 6.97 keV lines -- is clearly detected in all observations. In addition, a \FeH\ Ly$\beta$ line seems to be present in two of the three observations.
In this Section, the X-ray continuum and ionized iron lines are used to obtain a description of the energy distribution of \hd, and put constraints especially on the Fe abundance determination from these lines.

\begin{figure} \centering{ 
\includegraphics[bb=0.5cm 5.5cm 20.6cm 25cm,clip=true,width=9cm]{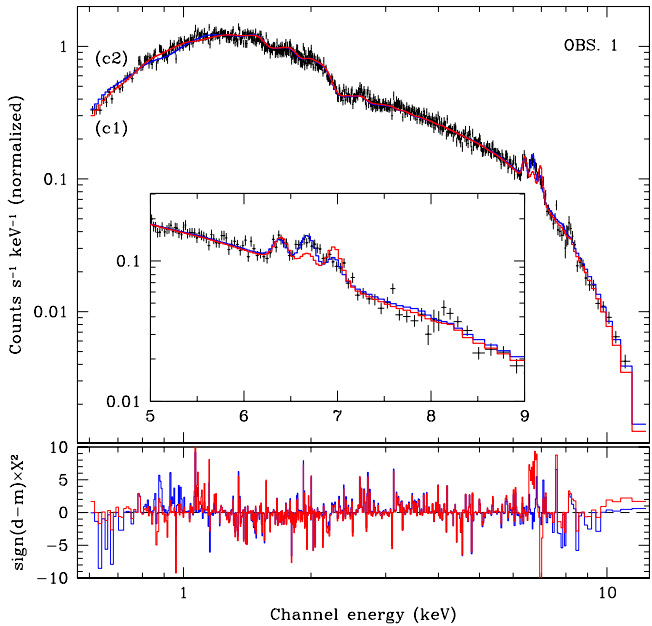}
\caption{The X-ray spectrum from \obsA\ showing the bad fit quality achieved in the framework of 1-T and 2-T models. Similar results are obtained for \obsB\ and \obsC). Curve 1 (red): N$_{\rm H}*$(T1+T2+GL); curve 2 (blue): N$_{\rm H_a}*$(T1)+N$_{\rm H_b}*$(T2+GL); (see text for details). The 1-T model fit overlaps curve 1 for E $>$ 1 keV and curve 2 for E $<$ 1 keV. In all cases a gaussian line is included at 6.4 keV. Note the poor fit of the 8.2 keV feature by the \textsc{mekal} model, which is similar to all tested configurations.} 
\label{fig:models2T}  } \end{figure}

Models with one or two thermal components do not give acceptable results. Replacing one of the thermal components by a {\it power law}, even though including a high energy cut-off, does not improve the quality of the fits. These models do not account simultaneously for the observed excess in the hard X-ray continuum and intensities of the He and H like
iron lines, or for the soft continuum, even if each component is affected by different absorption columns. Likewise, the fits are not much improved by the addition of a second absorption for the hard component.


Thus, we note that the \XMM\ spectrum strongly contrasts with the 1-T model derived by \citet{TO01} from BeppoSAX observations. For example, we show in Fig. \ref{fig:models2T} the spectrum from \obsA\ and the 1-T and 2-T fits discussed above. A 1-T model results in N$_{\rm H}$ $\sim$ 0.3$\times$10$^{22}$ cm$^{-2}$, $k$T $\sim$ 15.2 keV, $Z$ $\sim$ 0.4 \Zsol, and $\chi^{2}_{\nu}$ = 1.3. 
For the 2-T model, Fig. \ref{fig:models2T} shows the case in which both thermal components are affected by the same absorption column [N$_{\rm H}*$(T1+T2+GL)], and the case in which each plasma is affected by different absorptions [N$_{\rm H_a}*$(T1)+N$_{\rm H_b}*$(T2+GL)]. In the first case, we derive N$_{\rm H}$ $\sim$ 0.33$\times$10$^{22}$ cm$^{-2}$, $k$T1 and $k$T2 equal to $\sim$ 0.76 and $\sim$ 15.13 keV, respectively, $Z$ $\sim$ 0.43 \Zsol, and $\chi^{2}_{\nu}$ = 1.23. For the second case, N$_{\rm H_a}$ and $k$T1 are equal to $\sim$ 0.34$\times$10$^{22}$ cm$^{-2}$ and 5.89 keV, N$_{\rm H_b}$ and $k$T2 of $\sim$ 0.28$\times$10$^{22}$ cm$^{-2}$ and 28.94 keV, respectively, $Z$ $\sim$ 0.26 \Zsol, and $\chi^{2}_{\nu}$ = 1.15. (For the 2-T models the abundance values refer to the hottest component; solar values for the others).

\begin{figure} \centering{ 
\includegraphics[bb=0.5cm 5.5cm 20.6cm 25cm,clip=true,width=9cm]{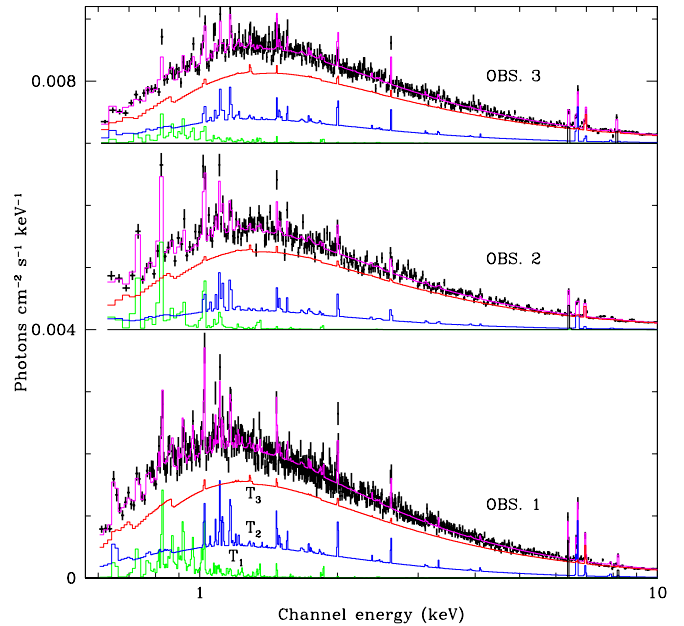}
\caption{Unfolded X-ray spectra at different epochs, added to constant values for clarity (as M1 in Table \ref{tbl:avrg_par}). 
} 
\label{fig:Xspec}  } \end{figure}

\begin{table*}[!] \caption{Best-fit parameters for the X-ray spectrum of \hd\ in each \XMM\ observation from a 3-thermal model.}              
\label{tbl:avrg_par}        \centering           \begin{tabular}{r c c c c c c c c c c c c c c c c c c}            

\hline                    
\hline\\[-2.2ex]

& \multicolumn{2}{c}{\obsA} &&& \obsB &&& \multicolumn{2}{c}{\obsC} \\
\cline{2-3} \cline{5-7} \cline{9-10}\\[-2ex]
& M1 & M1d && M1 & [M1] & M1d && M1 & M1d \\
\hline\\[-2.2ex]

N$_\mathrm{H_{a}}$ (10$^{22}$\,cm$^{-2}$)    	 	& 0.34$^{+0.01}_{-0.01}$   	& 0.96$^{+0.11}_{-0.14}$       	 && 0.42$^{+0.05}_{-0.04}$		  &  [0.46$^{+0.04}_{-0.04}$]	  & 0.80$^{+0.22}_{-0.23}$ && 0.44$^{+0.02}_{-0.02}$	  	 & 1.32$^{+0.28}_{-0.67}$\\[0.2ex]
$k$T$_{1}$ (keV)				    	& 0.66$^{+0.05}_{-0.05}$       & 0.23$^{+0.04}_{-0.03}$         	 && 0.37$^{+0.07}_{-0.15}$		  &  [0.30$^{+0.07}_{-0.04}$]	  & 0.36$^{+0.18}_{-0.09}$ && 0.66$^{+0.26}_{-0.09}$	& 0.21$^{+0.12}_{-0.04}$  \\[0.2ex]
$f_{{\rm T}_{1}}$ (erg\,cm$^{-2}$\,s$^{-1}$)	    	& 7.8$\times$10$^{-13}$	   	   & 4.6$\times$10$^{-11}$		                	 &&  1.1$\times$10$^{-12}$	       &  [2.7$\times$10$^{-12}$]	& 6.3$\times$10$^{-12}$&&  5.2$\times$10$^{-13}$	    &	7.1$\times$10$^{-11}$			\\[0.2ex]
EM$_{\rm T_{1}}$ (10$^{55}$\,cm$^{-3}$)	    	 		& 0.03 		       & 2.1			        		 &&  0.04		       & [0.1]  			&0.3&&   0.02 			& 3.4\\[0.2ex]
N$_\mathrm{H_{b}}$ (10$^{22}$\,cm$^{-2}$)  		& ...			       & 0.23$^{+0.09}_{-0.05}$               && ...					      & [...]		      &  0.17$^{+0.18}_{-0.09}$ && ...				 & 0.97$^{+0.53}_{-0.60}$     \\[0.2ex]
$k$T$_{2}$ (keV)				    	& 5.40$^{+0.60}_{-0.81}$       & 5.90$^{+1.72}_{-1.05}$         	 && 3.49$^{+1.12}_{-0.91}$		  &  [1.55$^{+0.37}_{-0.22}$]	 & 4.69$^{+2.33}_{-1.28}$ && 5.68$^{+3.80}_{-1.01}$	  & 4.45$^{+5.67}_{-1.20}$\\[0.2ex]
$f_{{\rm T}_{2}}$ (erg\,cm$^{-2}$\,s$^{-1}$)	    	& 9.1$\times$10$^{-12}$	   	       & 8.8$\times$10$^{-12}$	                	 &&  5.5$\times$10$^{-12}$	      &  [1.4$\times$10$^{-12}$]	 & 4.2$\times$10$^{-12}$ &&  7.9$\times$10$^{-12}$		& 7.4$\times$10$^{-12}$			\\[0.2ex]  
EM$_{\rm T_{2}}$ (10$^{55}$\,cm$^{-3}$)	    	 		& 0.5  		       &	0.4		        		 &&  0.3 			      & [0.09]  		&0.2	&&  0.4  		&0.4		\\[0.2ex]
N$_\mathrm{H_{c}}$ (10$^{22}$\,cm$^{-2}$) 		& ...			       & 0.47$^{+0.07}_{-0.07}$               && ...					      & [...]		      & 0.51$^{+0.08}_{-0.08}$ && ...				 & 0.38$^{+0.11}_{-0.03}$      \\[0.2ex]
$k$T$_{3}$ (keV)				    	& 27.17$^{+3.61}_{-4.01}$  	& 20.76$^{+3.21}_{-2.42}$       	 && 20.82$^{+4.67}_{-3.80}$		  &  [14.26$^{+1.95}_{-1.28}$]   & 16.87$^{+4.02}_{-2.46}$  && 36.86$^{+7.24}_{-7.68}$	  & 34.48$^{+13.68}_{-8.95}$\\[0.2ex]
$f_{{\rm T}_{3}}$ (erg\,cm$^{-2}$\,s$^{-1}$)		& 3.7$\times$10$^{-11}$	   	  & 3.8$\times$10$^{-11}$		                	 &&  3.2$\times$10$^{-11}$	      &  [3.7$\times$10$^{-11}$]	& 3.4$\times$10$^{-11}$ && 3.1$\times$10$^{-11}$		&	3.4$\times$10$^{-11}$		\\[0.2ex]  
EM$_{\rm T_{3}}$ (10$^{55}$\,cm$^{-3}$)	 			& 1.9  		       &	2.0		        		 &&  1.7 		      & [1.9]				& 1.7&&  1.7  			&	1.7	\\[0.2ex]

$Z_{\rm T_{3}}$ (\Zsol)				    	& 0.31$^{+0.12}_{-0.09}$       & 0.24$^{+0.10}_{-0.14}$         	 && 0.45$^{+0.13}_{-0.12}$		  &  [0.44$^{+0.08}_{-0.08}$]	  & 0.37$^{+0.11}_{-0.13}$ && 0.52$^{+0.18}_{-0.17}$	  & $<$ 0.89\\[0.2ex]

Line (keV)				     		& 6.39$^{+0.04}_{-0.01}$     & 6.40$^{+0.02}_{-0.02}$	        	 &&  6.41$^{+0.02}_{-0.02}$		&  [6.41$^{+0.02}_{-0.02}$]	& 6.41$^{+0.02}_{-0.02}$ &&  6.4$^{\mathrm{b}}$		&  6.4$^{\mathrm{b}}$	\\[0.2ex]
$\sigma_{\rm Line}$ (keV) 		  		& $<$ 0.05 	       	     & 0.01$^{\mathrm{b}}$              	 &&  0.01$^{\mathrm{b}}$ 	&  [0.01$^{\mathrm{b}}$]	&  0.01$^{\mathrm{b}}$ && 0.01$^{\mathrm{b}}$  		 & $<$ 0.10\\[0.2ex]
Line (keV)				   		& 8.2$^{\mathrm{b}}$       & 8.2$^{\mathrm{b}}$                 	 &&  ... 			&  [...]  			& ... &&  8.16$^{+0.08}_{-0.06}$		 &  8.2$^{\mathrm{b}}$\\[0.2ex]
$\sigma_{\rm Line}$ (keV) 		   		& $<$ 0.11      	    & 0.001$^{\mathrm{b}}$              	 &&  ... 			&  [...]  			& ... &&  $<$ 0.12				&  0.001$^{\mathrm{b}}$ \\[0.2ex]
					
$f_{tot}$ (erg\,cm$^{-2}$\,s$^{-1}$) 			& 4.8$\times$10$^{-11}$	       & 9.3$\times$10$^{-11}$		                	 && 3.9$\times$10$^{-11}$	      & 	     [4.1$\times$10$^{-11}$]	& 4.4$\times$10$^{-11}$ && 4.2$\times$10$^{-11}$      & 1.1$\times$10$^{-10}$		\\[0.2ex]  
$\chi^{2}_{\nu}$/d.o.f.$^{\mathrm{a}}$	     		& 1.05/588		   	& 1.04/588		        	 && 1.13/270		  & [1.13/270]  		& 1.12/268  && 1.00/386			  		& 1.00/385\\[0.2ex]
\hline
                                  \end{tabular} 
\begin{list}{}{}
\item M1: \textsc{N$_{\mathrm{H_a}}*$(T$_1$+T$_2$+T$_3$+2 g.\,lines)}; M1d: \textsc{N$_{\mathrm{H_a}}*$T$_1$+N$_{\mathrm{H_b}}*$T$_2$+N$_{\mathrm{H_c}}*$(T$_3$+2 g.\,lines)}. 
\item $^{\mathrm{a}}$ degrees of freedom; $^{\mathrm{b}}$ frozen parameter.
\item Notes: solar abundances for T$_1$ and T$_2$. 
Fluxes are given unabsorbed in the 0.2--12 keV energy band.
Quoted errors are at the 90\% confidence level.

\end{list}
\end{table*}

X-ray spectra are better described by the sum of three thermal components (3-T): a {\it cool} ($k$T$_{1}$ $\sim$ 0.4--0.7 keV), a {\it warm} ($k$T$_{2}$ $\sim$ 3--6 keV), and a {\it hot} plasma ($k$T$_{3}$ $\sim$ 21--37 keV; see M1 in Table \ref{tbl:avrg_par} and Fig. \ref{fig:Xspec}).
About 80\% of the 0.2--12 keV flux is due to the {\it hot} plasma, notably as for \gcas\ \citep{Smith04},  while the {\it warm} and {\it cool} plasmas account for the other $\sim$ 18\% and $\sim$ 2\%, respectively. The same distribution is followed by the emission measure of each thermal component, with respect to the total value.
The ionized iron lines are well described for this model, in which they arise from a combination of the {\it hot} ($k$T $\sim$ 20 keV) and {\it warm} ($k$T $\sim$ 4 keV) plasmas. 
The {\it hot} component accounts for a dominant contribution to the \FeHe\ ($\sim$ 58--72\%) and \FeH\ ($\sim$ 82--95\%) strength.
Therefore, the 
high temperature of the {\it hot} component is required by the shape of the continuum and  by the intensity of the ionized iron lines, being thus a reliable value even though derived from the limited 0.6--12 keV range. [In addition, it is worth noting that a significant fraction of the bolometric flux in X-rays ($\sim$ 50\%) of a plasma with $k$T $\sim$ 30 keV is emitted in this energy range.]
Finally, while a high temperature is otherwise uncertain for the EPIC instruments, at least the high temperature in the solution of \obsA\ and \obsC\ is correlated with the presence of the Ly$\beta$ line at 8.2 keV.

The Ly$\beta$ line, suspected in \obsA\ and \obsC, cannot be reproduced by any thin thermal \textsc{mekal}\footnote{http://heasarc.nasa.gov/xanadu/xspec/manual/manual.html} model we have tested, and a gaussian line was added. A second gaussian line accounts for the Fe fluorescence line at 6.4 keV, not included in the \textsc{mekal} code.

The column density derived from M1, $\sim$ 3--4$\times$10$^{21}$\,cm$^{-2}$, is larger than that due to the interstellar medium \citep[$\sim$ 2$\times$10$^{21}$\,cm$^{-2}$;][]{Rachford01}. 
We note that these values
are much lower than the values ($\sim$ 2--5$\times$10$^{22}$\,cm$^{-2}$) derived by \citet{SB06}, but these latter were
determined from emission V/R peaks of partially optically thick
lines for formation regions coinciding with orbital disk segments aligned
along the line of sight, thereby incurring  
absorption columns that are likely to be several
times larger than the column to the Be star itself.\footnote{A N$_{\rm H}$ $\ga$ 10$^{22}$\,cm$^{-2}$ fails to produce satisfactory fits to the fluxes below $\sim$ 1.5 keV in M1.}
We hope to address the columns, and hence the geometry of the X-ray sites with a firmer column estimate from a planned high resolution Reflection Grating Spectrometer (RGS) observation along with medium resolution (EPIC) \XMM\ spectra -- scheduled for early September, 2007.

In \obsA\ and \obsC\ the fitting process following the model M1 converges toward a
unique solution. In \obsB, however, the fitting process finds two equally
probable configurations with quite different temperatures for the three components, as well as different flux contributions and emission measures for the {\it cool} and {\it warm} components (Table 
\ref{tbl:avrg_par}).  
It seems, however, that the models with the highest temperatures provide a better
description of the iron line complex. In the ``cooler convergence" for \obsB, the {\it cool} and {\it warm} plasma account for 4\% each of the total value, and $\sim$ 6.5\% and $\sim$ 3.5\%, respectively, of the total 0.2--12 keV flux -- in contrast with those contributions of the ``hotter convergence" for \obsB\ and values derived from the other observations, summarized above.  

Contrary to the {\it warm} and {\it cool} components, in which the fits result in abundances which are consistent with solar values for all elements,  sub-solar abundances are clearly needed for the model parameters of the {\it hot} component. Using the \textsc{vmekal} model, we checked that setting a lower Fe abundance for the {\it hot} component while keeping the abundances of other metals to their solar values gives results consistent with those obtained by freeing the elemental abundances altogether ($Z \la 0.5Z_{\odot}$). 
Thus, this value is determined from the EW of the ionized iron lines.

So distinct temperatures strongly suggest complex X-ray environments. In this sense, we tried a number of 3-T models in which we: 
\begin{description}
\item (a) - determined two absorption columns, one for the {\it hot} component
  and another for the {\it warm} \& {\it cool} components; 
\item (b) - divided the {\it hot} plasma in (a) into two unequal
subcomponents. The absorption of the first subcomponent is determined
independently of the absorption of the other subcomponent, which is
tied to the {\it warm} and {\it cool} components.
The relative fractions of the two {\it hot} components is determined as a
free parameter;
\item (c) - repeated the (b) model but forced the emission measures
  of the two {\it hot} subcomponents to be equal;
\item (d) - determined three absorption columns, one for each thermal component. 
\end{description}
 
Models (a,b,c) do not improve the fits obtained from M1, and with two exceptions the derived
parameters are consistent with those of M1 at the 90\%
confidence level.
The two exceptions concern the {\it hot} component and its absorption following the model (b). According to this model, $\sim$ 15--50\% of the hottest component is more strongly absorbed ($\sim$ 10$\times$) than the other $\sim$ 50--85\%. The latter {\it hot} subcomponent is tied to the absorption column of the {\it warm} and {\it cool} components, and the parameters for this column are consistent with those derived from M1 (see Table \ref{tbl:avrg_par}).
Therefore, we can speculate that, similar to the X-ray spectrum of \gcas\ \citep{Smith04}, the hottest component of \hd\ is affected by multiple absorption columns.
As a second exception, the {\it hot} temperatures in each observation are systematically cooler than those of M1 by about 30\%.
We also note that the (a,b,c) models result in two convergences for \obsB. Both are  consistent with those obtained from M1 in Table \ref{tbl:avrg_par}.
The results of the models briefly discussed above corroborate the complex nature of the X-ray spectrum of \hd, already inferred from the simplified model M1.

Model (d) is somewhat different from those discussed above (see Table \ref{tbl:avrg_par}). First, the total unabsorbed flux is systematically higher in each observation, notably for \obsA\ and \obsC. Second, the {\it cool} component, except for \obsB, is now even cooler. This component is strongly absorbed and accounts for a major fraction of the total flux (ranging from 14\% to 64\% for different observations). Also, the {\it cool} plasma has a larger emission measure than the M1 model.
No significant difference arises for fluxes and emission measures of the {\it hot} and {\it warm} components. We also note that this model results in a slightly lower temperature for the {\it hot} component.

As an alternative to the 3-T models discussed above, a good description of the observed X-ray energy distribution of \hd\ is also obtained using a model composed of two thermal components (a {\it cool}, $k$T $\sim$ 0.2--1.6 keV, and a {\it hot}, $k$T $\sim$ 8--12 keV) added to a {\it power law} with a hard photon index (see Table \ref{tbl:avrg_par_b}).  As in the 3-T models, the \FeHe\ and \FeH\ lines are well fitted by the thermal plasma model, and subsolar abundances, at least for Fe, are needed. In the 2-T + PL model, the {\it power law} accounts for part of the hard X-ray continuum ($\sim$ 20\% of the total 0.2--12 keV flux in all observations) and the iron lines can then be well represented by a single thermal component of temperature ($k$T $\sim$ 10 keV) intermediate between those of the {\it warm} and {\it hot} plasmas in the 3-T model, and consistent with that obtained by \citet{TO01}. The flux contributions of the {\it hot} plasma is about $\sim$ 55\,\% of the total 0.2--12 keV flux in \obsA, $\sim$ 70\% in \obsB, and $\sim$ 40\% in \obsC. However, as supported by the presence of the Ly$\beta$ line, we believe that the {\it power law} does not give a realistic description of the high energy tail in \hd.

\begin{table} \caption{Best-fit parameters for the X-ray spectrum of \hd\ in each \XMM\ observation, from thermal models added to a {\it power law} component.}              
\label{tbl:avrg_par_b}        \centering           \begin{tabular}{r c c c c c c c c c c c c c c c}            

\hline                    
\hline\\[-2.2ex]

& \obsA & \obsB & \obsC \\

\hline\\[-2.2ex]
\multicolumn{4}{l}{M2: \textsc{N$_{\rm Ha}$}$*$\textsc{T$_1$+N$_{\rm Hb}*$(T$_2$+p.\,law+2 g.\,lines)}} \\[0.2ex]

N$_{\rm Ha}$ (10$^{22}$\,cm$^{-2}$)	 	& 	0.92$^{+0.31}_{-0.31}$           & 	     0.74$^{+0.19}_{-0.19}$             &	      1.10$^{+1.29}_{-0.62}$		\\[0.2ex]
$k$T$_{1}$ (keV)				 & 	0.22$^{+0.04}_{-0.03}$        &       0.35$^{+0.13}_{-0.10}$          &	      0.21$^{+0.16}_{-0.10}$		\\[0.2ex]
$f_{{\rm T}_{1}}$ (erg\,cm$^{-2}$\,s$^{-1}$)	    	& 3.9$\times$10$^{-11}$		   			&  4.1$\times$10$^{-12}$	&  2.7$\times$10$^{-11}$		          		\\[0.2ex]
EM$_{\rm T_{1}}$ (10$^{55}$\,cm$^{-3}$)	 & 	1.8     	              & 	      0.2		               &	      1.3			\\[0.2ex]
N$_{\rm Hb}$ (10$^{22}$\,cm$^{-2}$)	 & 	0.37$^{+0.03}_{-0.02}$           & 	      0.39$^{+0.04}_{-0.03}$             &	      0.44$^{+0.03}_{-0.03}$		\\[0.2ex]
$k$T$_{2}$ (keV)				 & 	8.76$^{+0.75}_{-0.82}$        & 	      10.24$^{+1.58}_{-1.46}$         &	      9.34$^{+1.42}_{-0.99}$		\\[0.2ex]
$f_{{\rm T}_{2}}$ (erg\,cm$^{-2}$\,s$^{-1}$)	    	& 3.2$\times$10$^{-11}$		   			&  3.0$\times$10$^{-11}$	&  2.6$\times$10$^{-11}$		          		\\[0.2ex]
EM$_{\rm T_{2}}$ (10$^{55}$\,cm$^{-3}$)	 & 	1.7     	              & 	      1.6		               &	      1.4	  \\[0.2ex]
$Z_{\rm T_{2}}$ (\Zsol)			 		 & 	0.43$^{+0.05}_{-0.06}$        & 	      0.42$^{+0.17}_{-0.10}$          &	      0.50$^{+0.39}_{-0.07}$		\\[0.2ex]
$\Gamma$					 & 	1.09$^{+0.15}_{-0.22}$        & 	      1.04$^{+0.79}_{-3.09}$          &	      1.02$^{+0.23}_{-1.03}$		\\[0.2ex]
$f_{{\rm \Gamma}}$ (erg\,cm$^{-2}$\,s$^{-1}$)	    	& 1.5$\times$10$^{-11}$		   			&  8.0$\times$10$^{-12}$	&  1.5$\times$10$^{-11}$		          		\\[0.2ex]
Line (keV)					 & 	6.39$^{+0.03}_{-0.02}$        & 	      6.41$^{+0.02}_{-0.02}$          &	      6.4$^{\mathrm{b}}$				\\[0.2ex]
$\sigma_{\rm Line}$ (keV) 			 & 	$<$0.06     	              & 	      $<$0.07  	               &	      0.01$^{\mathrm{b}}$				\\[0.2ex]
Line (keV)					 & 	8.2$^{\mathrm{b}}$            & 	      ...		               &	      8.16$^{+0.07}_{-0.06}$		\\[0.2ex]
$\sigma_{\rm Line}$ (keV) 			 & 	 0.001$^{\mathrm{b}}$         & 	      ...		               &	      $<$0.11  			\\[0.2ex]

$f_{\rm tot}$ (erg\,cm$^{-2}$\,s$^{-1}$)		 & 	8.6$\times$10$^{-11}$         & 	      4.2$\times$10$^{-11}$           &	      6.8$\times$10$^{-11}$			\\[0.2ex]  
$\chi^{2}_{\nu}$/d.o.f.$^{\mathrm{a}}$		 & 	1.07/588     	              & 	      1.17/268 	               &	      1.01/385 			\\[0.2ex]

\hline\\[-2.2ex]
\multicolumn{4}{l}{M3: \textsc{N$_{\rm H}$}$*$\textsc{(cemekl+p.\,law+2 g.\,lines)}} \\[0.2ex]

N$_{\rm H}$ (10$^{22}$\,cm$^{-2}$)	 	 & 0.34$^{+0.01}_{-0.01}$   	&  0.38$^{+0.01}_{-0.02}$  	& 0.45$^{+0.01}_{-0.01}$	    	\\[0.2ex]
T$_{max}$ (keV)					 & 27.90$^{+7.90}_{-4.98}$   	&  35.20$^{+10.93}_{-8.33}$ & 29.51$^{+13.92}_{-6.45}$		\\[0.2ex]
$Z$ (\Zsol)			 		 & 0.53$^{+0.10}_{-0.08}$   	&  0.55$^{+0.11}_{-0.06}$  	& 0.58$^{+1.56}_{-0.09}$ 	    	\\[0.2ex]
$f_{{\rm T}}$ (erg\,cm$^{-2}$\,s$^{-1}$)	    	& 3.7$\times$10$^{-11}$		   			&  3.6$\times$10$^{-11}$	&  3.2$\times$10$^{-11}$		          		\\[0.2ex]
$\Gamma$					 & 1.13$^{+1.79}_{-0.23}$	
&  1.1$^{\mathrm{b}}$  			& 1.04$^{+2.48}_{-0.76}$		\\[0.2ex]
$f_{{\rm \Gamma}}$ (erg\,cm$^{-2}$\,s$^{-1}$)	    	& 1.1$\times$10$^{-11}$		   			&  3.0$\times$10$^{-12}$	&  1.1$\times$10$^{-11}$		          		\\[0.2ex]
Line (keV)					 & 6.4$^{\mathrm{b}}$   		&  6.41$^{+0.02}_{-0.02}$  	& 6.4$^{\mathrm{b}}$			    	\\[0.2ex]
$\sigma_{\rm Line}$ (keV) 			 & 0.01$^{\mathrm{b}}$   		&  $<$ 0.07  		& 0.01$^{\mathrm{b}}$			    	\\[0.2ex]
Line (keV)					 & 8.20$^{+0.12}_{-0.09}$   	&  ...  			& 8.16$^{+0.06}_{-0.05}$    	\\[0.2ex]
$\sigma_{\rm Line}$ (keV) 			 &  0.001$^{\mathrm{b}}$   	
&  ...  			&  0.001$^{\mathrm{b}}$			    	\\[0.2ex]
$f_{\rm tot}$ (erg\,cm$^{-2}$\,s$^{-1}$)	 & 4.8$\times$10$^{-11}$   		&  3.9$\times$10$^{-11}$   	& 4.3$\times$10$^{-11}$     	\\[0.2ex]  
$\chi^{2}_{\nu}$/d.o.f.$^{\mathrm{a}}$		 		 & 1.12/592   			&  1.17/272 		& 1.05/389			    	\\[0.2ex]

\hline                    
                                  \end{tabular} 
\begin{list}{}{}
\item $^{\mathrm{a}}$ degrees of freedom; $^{\mathrm{b}}$ frozen parameter.
\item Notes: 
M2: solar abundances for T$_1$. M3: $\alpha$ = 1.
Fluxes are given unabsorbed in the 0.2--12 keV energy band.
Quoted errors are at the 90\% confidence level.
\end{list}
\end{table}

We tried the \textsc{cemekl}\footnote{A multi-temperature emission model built from the \textsc{mekal} code -- emission continuum/line spectrum from hot diffuse gas --, in which the emission measures of the plasmas follow a power law in temperature:  EM $\propto$ (T/T$_{max}$)$^\alpha$.} model, largely used in cataclysmic variables systems (CVs) leaving the $\alpha$ parameter free. The fit converges towards $\alpha$ values of 1, in agreement with the adiabatic case. The resulting temperatures $k$T$_{max}$ in the three observations range from 40 to 60 keV. 
However, this model gives slightly larger $\chi^{2}_{\nu}$ than those of models discussed above for two observations ($\chi^{2}_{\nu}$=1.14,  1.15, and 1.06, for \obsA, {\small 2}, and {\small 3}, respectively). It also predicts \FeH\ and \FeHe\ lines stronger and weaker than observed, respectively, and on this basis can probably be excluded. 
The bad fit quality achieved by the \textsc{cemekl} model to describe the ionized \FeHe\ and \FeH\ lines indicates that the 3-T description is most likely not the approximation of a spectrum emitted by a single region with a continuously variable temperature distribution, but rather reflects the existence of three physically separated sites of emission.

The addition of a {\it power law} component to the \textsc{cemekl} model improves the fits -- due to the better description of the high energy part of the spectrum -- but it does not provide mean $\chi^{2}_{\nu}$ values as low as those given by the 3-T model (see Table \ref{tbl:avrg_par_b}). We also tried a \textsc{cemekl} + soft thermal component to test the idea that part of the softest emission could arise in the shocked wind of the early type star. Again, the fit is improved but remains significantly worse than that of the 3-T model, primarily because of its inability to represent the details of the iron complex emission. 

From the 3-T model shown in Table \ref{tbl:avrg_par} the unabsorbed flux in the 2--10 keV energy band is $\sim$
2.3--2.9$\times$10$^{-11}$\,erg\,cm$^{-2}$\,s$^{-1}$ in the \XMM\ observations. This value is slightly lower than those obtained by BeppoSAX \citep[$\sim$ 3.2$\times$10$^{-11}$\,erg\,cm$^{-2}$\,s$^{-1}$;][]{TO01} and HEAO-1
\citep[$\sim$ 4.7$\times$10$^{-11}$\,erg\,cm$^{-2}$\,s$^{-1}$;][]{Tuohy88} satellites. Assuming a distance of $d$ = 300 pc to \hd\ \citep{Perryman97}, implies a luminosity of $\sim$ 2.6--3.1\,$\times$10$^{32}$\,\ergs. In the 0.2--12 keV band, the
unabsorbed \XMM\ flux is about 3.9--4.8                              $\times$10$^{-11}$\,erg\,cm$^{-2}$\,s$^{-1}$, or $L_{x}$ $\sim$ 4.2--5.2\,$\times$10$^{32}$\,erg\,s$^{-1}$ for $d$ = 300 pc (Table \ref{tbl:avrg_par}).

In conclusion, our analysis strongly suggest that the X-ray emission of \hd\ is a composition of three discrete plasmas with different volumes and temperatures. 
The highly ionized iron lines supports a thermal nature for the X-ray emission, and the presence of a very hot thermal component.
However, we cannot conclusively argue if each component is affected by distinct absorptions or not. If not, then the {\it hot} plasma dominates the total X-ray emission ($\sim$ 80\%). If so, the coolest plasma is most likely strongly absorbed and responsible for a contribution in flux similar to the flux due to the hottest plasma.

\begin{figure} \centering{ 
\includegraphics[bb=0.5cm 5.5cm 20.6cm 25cm,clip=true,width=9cm]{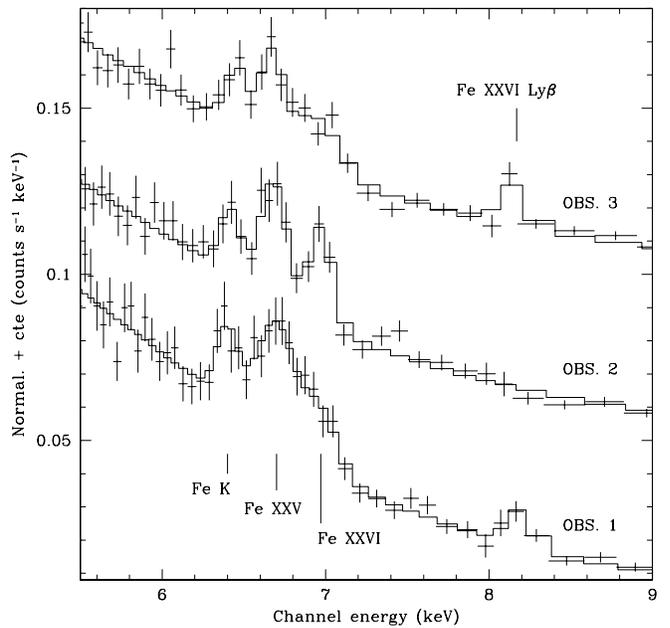}
\caption{The Fe\,K$\alpha$ complex and the suspected \FeH\ Ly$\beta$ line of \hd\ seen in different epochs. The $pn$ single-spectra are shown normalized and added to a constant term for clarity. See EWs in Table \ref{tbl:EW_FeKa}.} 
\label{fig:fekalpha}} \end{figure}

\begin{table*} \caption{Parameters of the iron emission lines detected in X-rays.}              
\label{tbl:EW_FeKa}        \centering           \begin{tabular}{l c c c c c c c c c c c c c c c c}            
\hline                    
\hline\\[-2.2ex]

 & \multicolumn{3}{c}{\obsA} && \multicolumn{3}{c}{\obsB} && \multicolumn{3}{c}{\obsC}	\\
\cline{2-4} \cline{6-8} \cline{10-12}\\[-2ex]
 & E$_{\rm C}$ & EW & Flux$^{\mathrm{a}}$ && E$_{\rm C}$ & EW & Flux$^{\mathrm{a}}$ && E$_{\rm C}$ & EW & Flux$^{\mathrm{a}}$ \\ 
 & (keV) & (eV) & ($\times$10$^{-5}$) && (keV) & (eV) & ($\times$10$^{-5}$) && (keV) & (eV) & ($\times$10$^{-5}$) \\ 

\hline
\noalign{\smallskip}

Fe\,K			& 6.40$^{+0.01}_{-0.04}$ 			& 47.8$^{+11.9}_{-14.0}$		& 1.9$^{+0.5}_{-0.5}$	&& 6.42$^{+0.02}_{-0.02}$ 	& 48.7$^{+16.3}_{-11.4}$	& 1.5$^{+0.5}_{-0.3}$	&& 6.46$^{+0.01}_{-0.03}$ & 54.2$^{+11.8}_{-12.1}$	& 1.9$^{+0.4}_{-0.4}$\\	[0.2ex]	
\FeHe 			& 6.70$^{\mathrm{b}}$				& 159.0$^{+34.3}_{-20.2}$ 		& 5.8$^{+1.2}_{-0.7}$	&& 6.68$^{+0.01}_{-0.01}$	& 126.0$^{+18.4}_{-16.2}$ 	& 3.7$^{+0.5}_{-0.5}$	&& 6.67$^{+0.05}_{-0.03}$ & 86.9$^{+88.0}_{-21.7}$	& 3.0$^{+3.1}_{-0.7}$\\	[0.2ex]   		
\FeH\ Ly$\alpha$   	& 6.97$^{\mathrm{b}}$				& 60.8$^{+19.0}_{-23.8}$		& 2.1$^{+0.7}_{-0.8}$	&& 6.97$^{+0.01}_{-0.01}$	& 99.1$^{+14.7}_{-13.6}$ 	& 2.6$^{+0.4}_{-0.4}$	&& 6.94$^{+0.11}_{-0.05}$ & 121.0$^{+40.1}_{-75.8}$	& 3.6$^{+1.2}_{-2.3}$\\	[0.2ex]	
\FeH\ Ly$\beta$ 	& 8.18$^{+0.03}_{-0.01}$ 			& 109.0$^{+41.2}_{-33.2}$ 		& 2.2$^{+0.8}_{-0.7}$	&& ...	& ...	& ...				&& 8.14$^{+0.02}_{-0.02}$ & 152.0$^{+42.8}_{-33.0}$	& 2.9$^{+0.8}_{-0.6}$\\	[0.2ex]	
	
\hline
                              \end{tabular} 
\begin{list}{}{}

\item [$^{\mathrm{a}}$] total flux in line, in units of photons\,cm$^{-2}$\,s$^{-1}$; $^{\mathrm{b}}$ frozen parameter.
\item Notes: EWs estimated from a \textsc{phabs$*$(bremss+3 gaussian lines)} model applied to spectra in the 5--10 keV energy range. E$_{\rm C}$ are the centroids of the Gaussian lines. Quoted errors are at 1\,$\sigma$.

\end{list}
\end{table*}

\subsection{The parameters of the iron lines in emission}
\label{section_fekalpha}

The lines of the \FeK\ complex are the strongest resolved emission lines in the 0.2--12 keV spectrum of \hd, and all are clearly present in each \XMM\ observation (Fig. \ref{fig:fekalpha}). 
In evaluating the parameters of the iron lines, a thermal bremsstrahlung model was used to describe the 5--10 keV continuum, and four Gaussian lines were added to represent each of the fluorescence, \FeHe, \FeH, and the suspected \FeH\ Ly$\beta$ lines.

Table \ref{tbl:EW_FeKa} lists the equivalent width of each Fe component in each observation.
The flux of the fluorescence line seems to be constant. 
On the other hand, the relative fluxes of each ionized component are slightly different in each epoch (Fig. \ref{fig:fekalpha} and Table \ref{tbl:EW_FeKa}), but their EWs are still mutually consistent at the $\pm$ 1 $\sigma$ level.
An apparent decrease is found for the He-like flux in each successive
observation, while the contrary is the case for the H-like ion.  

  The suspected presence of the \FeH\ Ly$\beta$ line at 8.2 keV is a novelty 
for spectra of any high-mass stellar X-ray source, whether emitted in the 
environs of the massive star or in a high-mass accretion binary system. 
Although the theory for the formation of this feature is not yet well
developed in the literature, one can make a few simple statements from 
what is known about the recombination spectrum of lighter hydrogenic ions. 
First, the Ly$\beta$ to Ly$\alpha$ strength ratio of such ions increases
with temperature, but rather only slowly so. For example, we have found 
that the
Ly$\beta$/Ly$\alpha$ line intensity ratio cannot be reproduced by any thin
thermal \textsc{mekal} model. We believe that the clear differences in the Ly$\beta$ strength noted in 
Fig. \ref{fig:fekalpha} between \obsB\ and \obsA\ cannot be explained
through the Boltzmann effect for the change in temperature, $\sim$ 21 keV
to 27 keV, respectively \citep[see e.g.][]{RSmith01}. Changes in this ratio are more likely caused
by a transition to a mildly optically thick regime, wherein the Ly$\alpha$
strength would be depressed through saturation. Alternatively, a suppressed
emission of Ly$\alpha$ could be caused by the superposition of some absorption
due to resonance scattering -- this would hint at a change in geometry of 
the plasma components. In either of the latter two alternatives, we might
expect to see a difference in the derived iron abundance in \obsB\ relative 
to the other two observations. However, as shown in Table \ref{tbl:FeK_cont}, 
any such effect must be small because the Fe abundances found for the 
three observations are internally consistent.  While we still believe either
of the latter two options offers a viable resolution of the problem, 
we must leave this as an unresolved issue.

\subsection{The iron lines in emission and the hard continuum}
\label{sec:feka_cont}

We applied the \textsc{cevmkl}\footnote{A \textsc{cemekl}-like model, in which the abundance of each element is allowed to vary.} model in the 6--8 keV energy range, in each observation, in order to estimate the ionization temperature ($k$T$_{ion}$) and iron abundance ($Z_{Fe}$) required by the \FeHe\ and \FeH\ lines. Using the same model in the 4--6 + 8.4--12 keV energy range, we also estimate the temperature ($k$T$_{h,cont}$) needed to describe only the hard continuum.

\begin{table} \caption{Temperatures from the Fe lines and hard X-ray continuum in the framework of 1-T model, and Fe abundance.}              
\label{tbl:FeK_cont}        \centering           \begin{tabular}{l c c c c c c c c c c c c}            
\hline                    
\hline\\[-2.2ex]

& $k$T$_{ion}$$^{\mathrm{a}}$ & $Z_{Fe}$$^{\mathrm{a}}$ & $k$T$_{h,cont}^{\mathrm{b}}$  \\
 & (keV) & ($\times$ solar)&  (keV) \\

\hline
\noalign{\smallskip}

\obsA & 8.74$^{+0.86}_{-0.78}$ & 0.23$^{+0.06}_{-0.02}$ & 17.76$^{+4.46}_{-4.22}$ \\[0.2ex]
\obsB & 10.68$^{+1.76}_{-1.63}$ & 0.27$^{+0.09}_{-0.08}$ & 15.29$^{+5.37}_{-3.56}$ \\[0.2ex]
\obsC & 9.87$^{+1.46}_{-1.06}$ & 0.26$^{+0.07}_{-0.06}$ & 24.62$^{+9.68}_{-4.89}$ \\[0.2ex]
\hline
                              \end{tabular} 
\begin{list}{}{}

\item [$^{\mathrm{a}}$] based in the \FeHe\ and \FeH\ lines, using the 6--8 keV energy range; $^{\mathrm{b}}$ from the 4--6 keV + 8.4--12 keV continuum.
\item Notes: Quoted errors are at the 90\% confidence level.
\end{list}

\end{table}

The results are shown in Table \ref{tbl:FeK_cont}.
This analysis strongly suggests that $k$T$_{h,cont}$ $>$ $k$T$_{ion}$, even though for \obsB\ at the 90\% confidence level, in agreement with the presence of the excess in the hard X-ray continuum. 
The iron abundance, in all observations, is consistent with 0.25 $\times$ $Z_{Fe,\odot}$.

\section{Timing behavior of the X-ray emission}

\begin{figure} \centering{ 
\includegraphics[bb=1cm 7.65cm 23.3cm 23cm,clip=true,width=10.2cm]{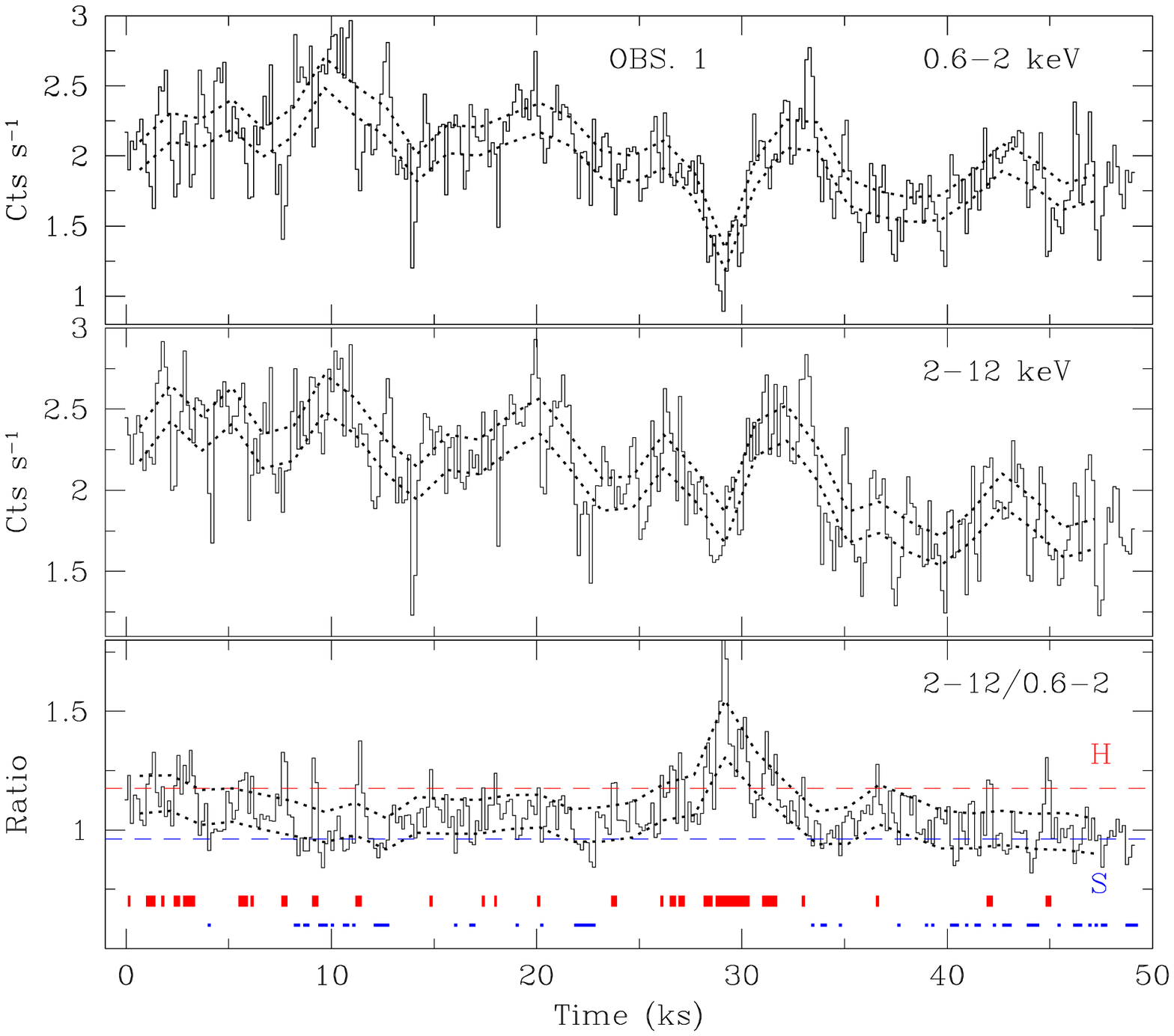}
\includegraphics[bb=1cm 7.65cm 23.3cm 22.9cm,clip=true,width=10.2cm]{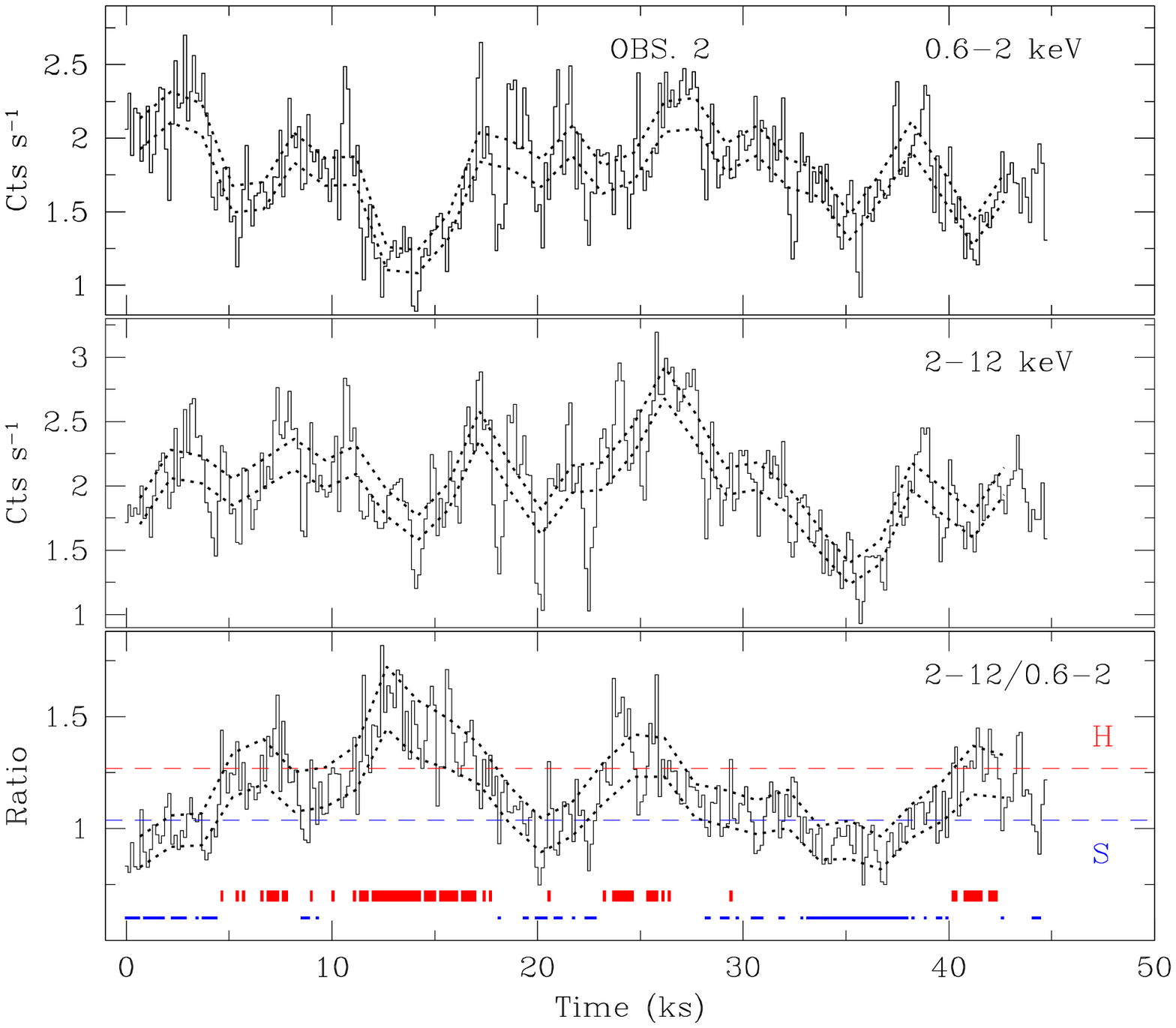}
\includegraphics[bb=1cm 5.9cm 23.3cm 22.9cm,clip=true,width=10.2cm]{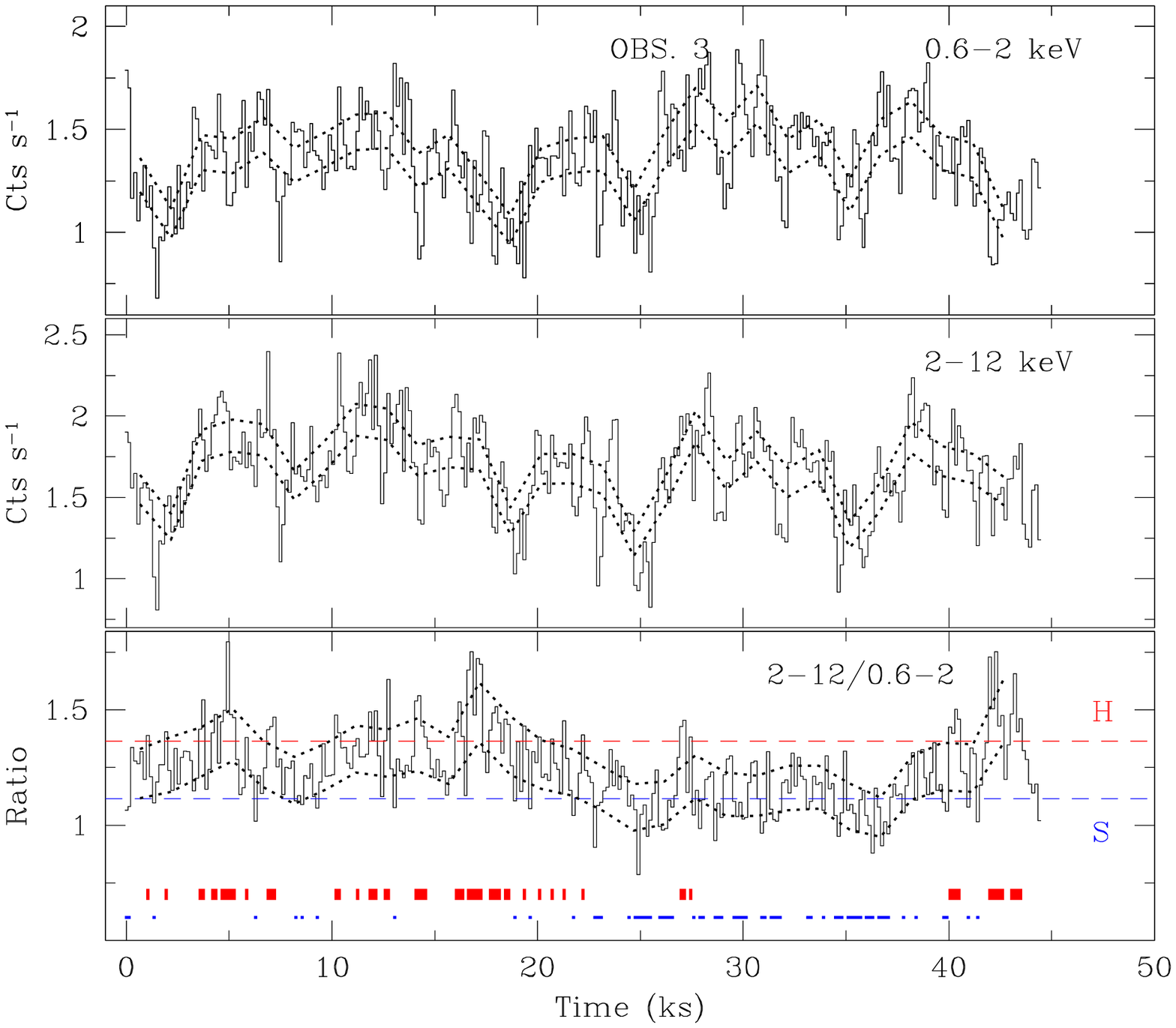}
\caption{EPIC light curves and hardness variabilities. Time bins of  150\,s. Dashed lines indicate the limits of the adopted hard (H) and soft
(S) states. The segment of lines in the bottom of the lower panels show the H and S time intervals. Dotted lines represent the $\pm$ 1\,$\sigma$ confidence belts (see text).
} 
\label{fig:lc_hrd}} \end{figure}

Figure \ref{fig:lc_hrd} shows background- and barycentric-corrected light curves of \hd\ in each XMM-{\it Newton} observation. We plot the light curves in the 0.6--2 keV and 2--12 keV energy bands, chosen to give comparable count rates, and the corresponding hardness ratio. 

At all epochs the source shows strong intensity and hardness variations, in which ubiquitous flare are observed. 
The dotted lines in Figure \ref{fig:lc_hrd} represents the 1\,$\sigma$ confidence belts derived from the photon statistic of 10 consecutive time bins of 150 s each. Interestingly, as for \gcas\ \citep{SRC98}, the shots in \hd\ are clearly superimposed over a slowly varying basal flux, and they are detected in time scales as short as 10 s, a limit determined by photon statistics.

There is no evidence for a hardness-intensity correlation in \hd. Even so, there are some interesting features. First, the average (absorbed) intensity seems to decrease over the subsequent observations (see Table \ref{tab:obslog}). (We note the divergent value for MOS2 on \obsB, most likely due to an inadequate for the off-axis position). Second, the dispersion of the hardness ratio distribution is more accentuated in the second and third observations (Fig. \ref{fig:hid}). Finally, rapid hardness variability reveals that the X-ray energy distribution varies on short time scales.

\subsection{Investigating the soft and hard states}

The strong and rapid variations, clearly observed in the light curves of \hd, are accompanied by a similar behaviour of the hardness ratios 
(hereafter HR; Figure \ref{fig:lc_hrd}). The changes are apparently random and occur on time scales as short as tens of seconds. 

\begin{figure*} \centering{ 
\includegraphics[bb=1.2cm 6.3cm 20cm 24.8cm,clip=true,width=6.338cm]{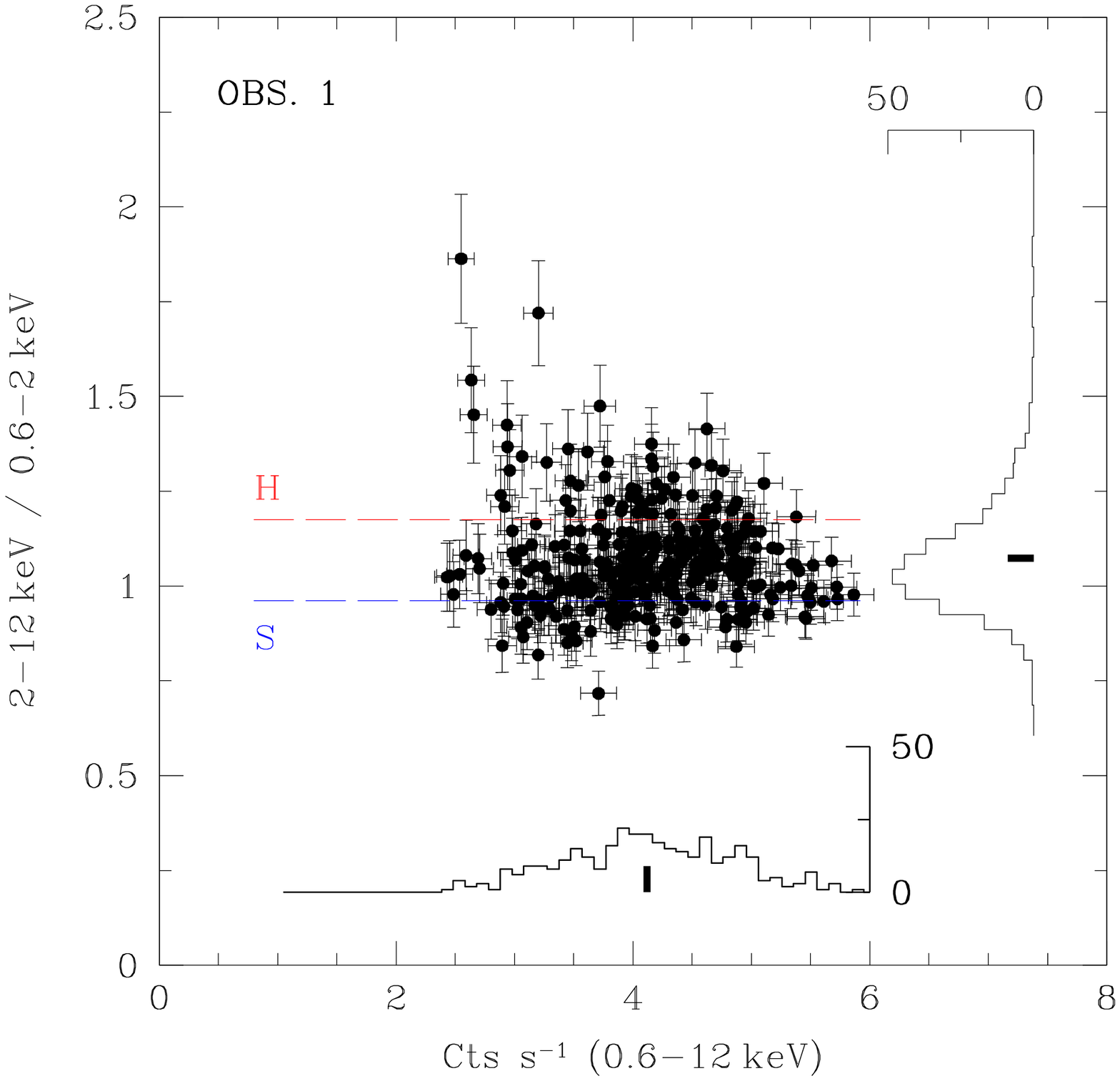}
\includegraphics[bb=3.65cm 6.3cm 20cm 24.8cm,clip=true,width=5.5cm]{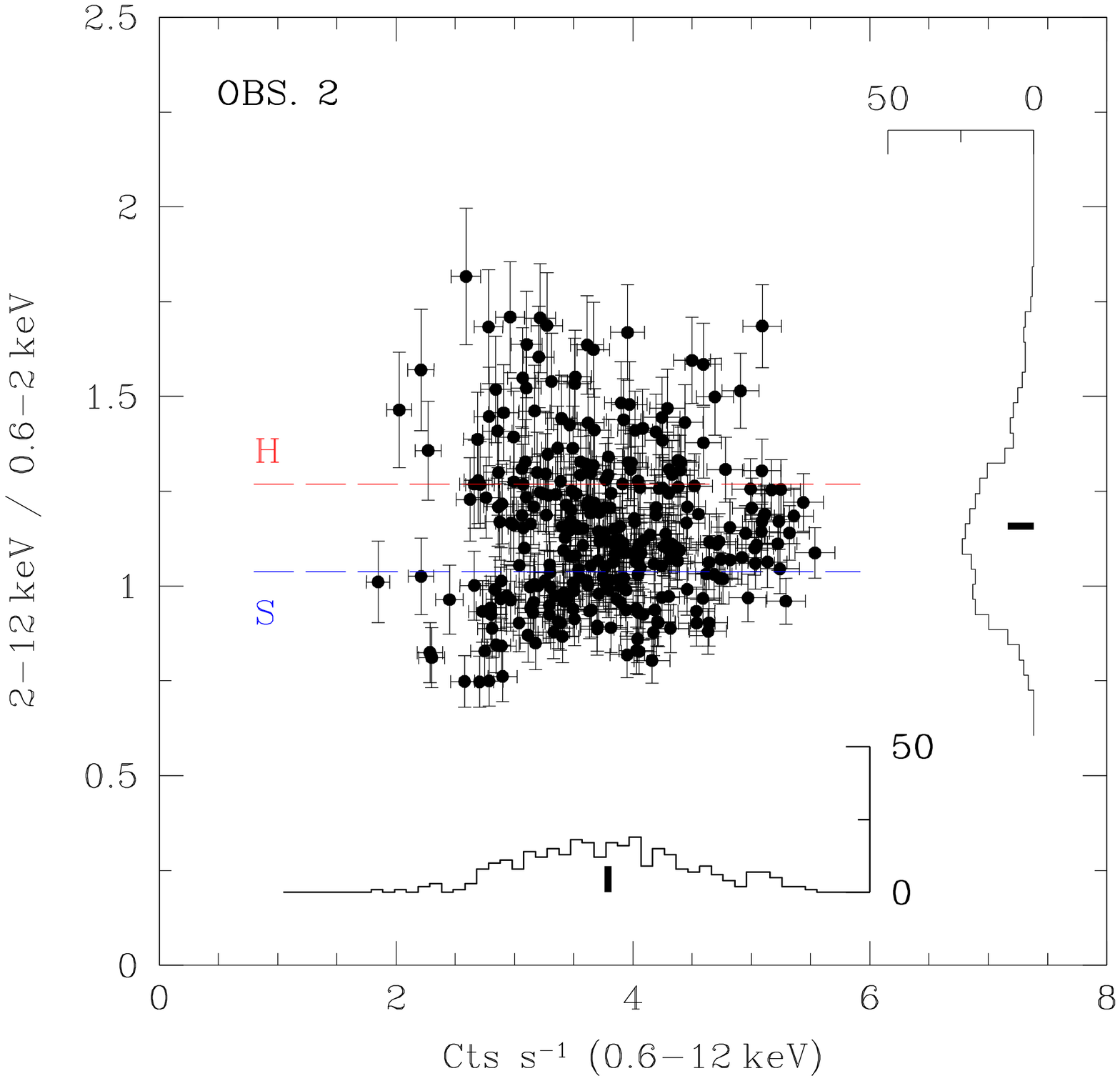}
\includegraphics[bb=3.65cm 6.3cm 20cm 24.8cm,clip=true,width=5.5cm]{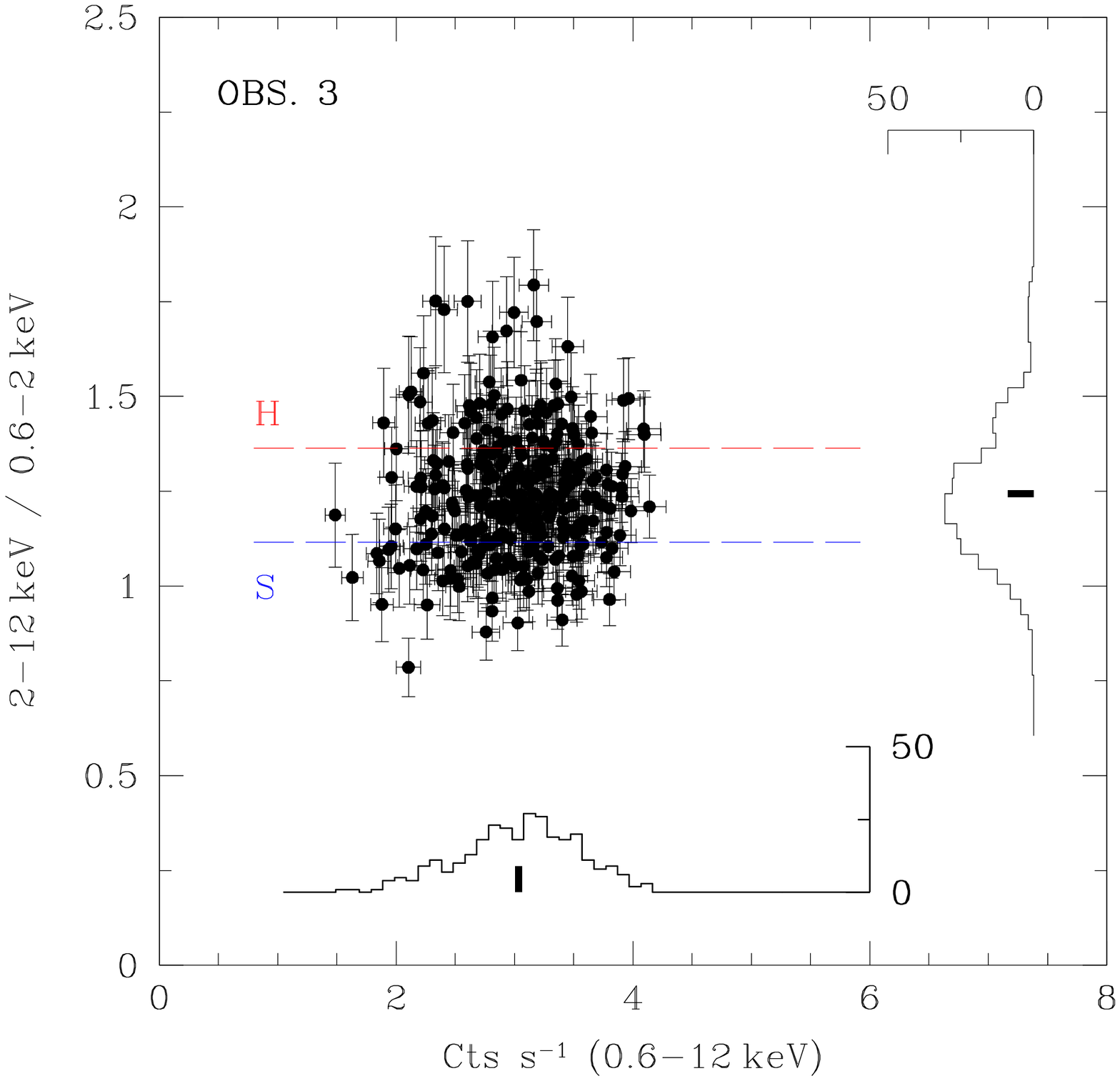}
\caption{Hardness-intensity X-ray diagrams. The insets show the intensities and hardness distributions, in which their mean values are marked. The data are grouped in time bins of 150\,s. The horizontal lines indicate the limit of the adopted hard and soft states. } 
\label{fig:hid}  } \end{figure*}

In order to try to investigate the detailed nature of the spectral variability of \hd, we went on to study spectra accumulated during times
of soft (S) and hard (H) dominated emission. 
The corresponding HR limits were adopted as being smaller than 0.9 and larger than 1.1 times
the mean HR for the ``soft" and ``hard" spectra, respectively. These limits are shown by dashed lines in Figures \ref{fig:lc_hrd} and \ref{fig:hid}. The times associated with the soft and hard states can be seen in the bottom of each lower panel of Figure \ref{fig:lc_hrd}. Times associated with enhanced solar-proton background were excluded from these
categories.

\begin{figure} \centering{ 
\includegraphics[bb=0.5cm 5.5cm 20.6cm 25cm,clip=true,width=9cm]{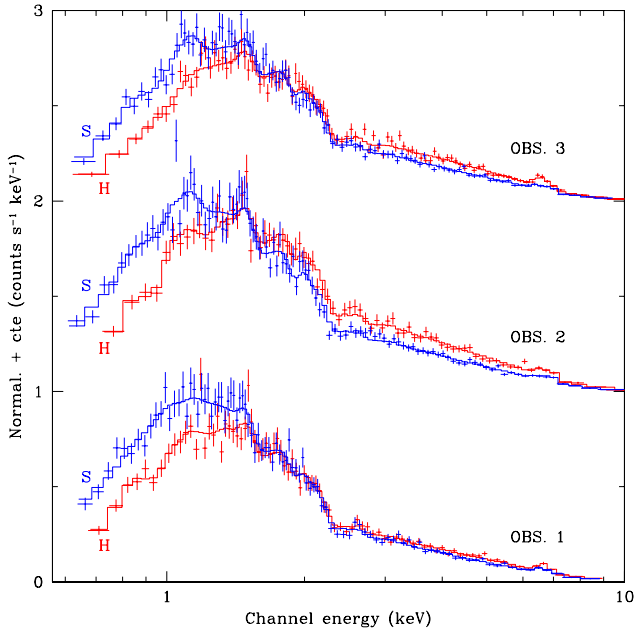}
\caption{X-ray spectra during high (H) and soft (S) states. Spectra are normalized and added to a constant term. 
The solid lines represent the resulting fits shown in Table \ref{tbl:spct_h_s_par}.} 
\label{fig:spct_h_s}} \end{figure}

Figure \ref{fig:spct_h_s} shows the counts spectra accumulated in the soft and hard states, and the resulting fits.
The best-fit parameters are shown in Table \ref{tbl:spct_h_s_par}. 
A soft photon deficit is clearly seen in the hard state and in
observations 2 and 3 it is accompanied by an increase of the flux of hard photons above $\sim$ 2 keV. In the framework of the 3-T model, it seems that most of the differences between the hard and the soft states are due to a change in the column density and/or to the luminosities of the {\it cool} and {\it warm} components. 
However, the temperatures of the {\it cool} and {\it warm} components are not statistically different in the hard and soft states and the direction of their evolution with state varies with the observation considered. 
The mean unabsorbed flux of the hottest temperature component is slightly ($\sim$ 10--40 \%) stronger in the hard states.

\subsection{Search for coherent pulsations}

We searched for coherent pulsations using the {\it PowSpec}/Xronos\footnote{http://heasarc.nasa.gov/docs/xanadu/xronos/xronos.html} v5.21, {\it Scargle}/Midas \citep{Scargle82}, and Z$^{2}_{n}$ \citep{Buccheri83} periodograms. The {\it PowSpec} and {\it Scargle} tools were applied on background-subtracted light curves. The Z$^{2}_{n}$ and also {\it PowSpec} were applied directly on the {\it pn} event lists.  
In all cases, we adopt the 0.6--2 keV, 2--12 keV, and 0.6--12 keV energy bands, and times corrected to the solar barycentric system.
The timing resolution of 200\,ms is set by the {\it pn} {\it extended full window} mode, corresponding to a Nyquist frequency of 2.5\,Hz. 

\begin{table*} 
\caption{Best-fit parameters of the X-ray spectra derived in soft (S) and hard (H) states, obtained from a M1-like model.}              
\label{tbl:spct_h_s_par}        \centering           \begin{tabular}{r c c c c c c c c c c }            

\hline                    
\hline\\[-2.2ex]                    
&\multicolumn{2}{c}{\obsA}	&&\multicolumn{2}{c}{\obsB}	&&\multicolumn{2}{c}{\obsC}\\
\cline{2-3} \cline{5-6} \cline{8-9}\\[-2ex]
& S & H && S & H && S & H\\
\hline                    
\noalign{\smallskip}

 N$_\mathrm{H}$ (10$^{22}$\,cm$^{-2}$)	 &     0.28$^{+0.04}_{-0.03}$  	   &     0.48$^{+0.07}_{-0.06}$       	  &&     0.30$^{+0.04}_{-0.03}$       	&	     0.73$^{+0.16}_{-0.13}$   &&     0.40$^{+0.04}_{-0.04}$  	    &	  0.60$^{+0.07}_{-0.06}$			       \\[0.2ex]
 $k$T$_{1}$ (keV)				 &     0.79$^{+0.24}_{-0.18}$  	   &     0.54$^{+0.09}_{-0.15}$       	  &&     0.65$^{+0.13}_{-0.12}$       	&	     0.30$^{+0.30}_{-0.10}$   &&     0.59$^{+0.15}_{-0.12}$  	    &	  0.62$^{+0.15}_{-0.22}$			       \\[0.2ex]
 $f_{\rm T_{1}}$ (erg\,cm$^{-2}$\,s$^{-1}$)	 &     7.3$\times$10$^{-13}$       &     2.2$\times$10$^{-12}$     	  &&     7.1$\times$10$^{-13}$	    	&	     1.1$\times$10$^{-11}$    &&     9.8$\times$10$^{-13}$	    &	  1.6$\times$10$^{-12}$		       \\[0.2ex] 
 $k$T$_{2}$ (keV)				 &  4.39$^{+1.41}_{-0.94}$    	   &     5.34$^{+2.27}_{-1.35}$       	  &&     2.97$^{+1.03}_{-0.68}$       	&	     2.52$^{+1.83}_{-0.87}$   &&     3.59$^{+1.02}_{-0.77}$  	    &	  5.96$^{+2.38}_{-1.33}$			       \\[0.2ex]
 $f_{\rm T_{2}}$ (erg\,cm$^{-2}$\,s$^{-1}$)	 &     9.5$\times$10$^{-12}$       &     1.2$\times$10$^{-11}$	   	  &&     3.6$\times$10$^{-12}$	 	&	     5.8$\times$10$^{-12}$    &&     6.9$\times$10$^{-12}$	    &	  8.8$\times$10$^{-12}$		       \\[0.2ex]	   
 $f_{\rm T_{3}}$ (erg\,cm$^{-2}$\,s$^{-1}$)	 &     3.4$\times$10$^{-11}$	   &     3.7$\times$10$^{-11}$            &&     2.8$\times$10$^{-11}$	   	&	     3.9$\times$10$^{-11}$    &&     3.0$\times$10$^{-11}$	    &	  3.6$\times$10$^{-11}$				       \\[0.2ex]
 $\chi^{2}_{\nu}$/d.o.f.				 &     1.07/88	        	   &	 0.98/78		   	  && 	1.05/98		    		&	     0.92/72		      &&     0.99/96			    &	  1.15/91			   \\[0.2ex]

\hline
                                  \end{tabular} 
\begin{list}{}{}
\item Notes: $f_{\rm T_{i}}$ is the unabsorbed 0.2--12 keV flux of the $k$T$_{i}$ component. The $k$T$_{3}$, abundances, and the central energy and dispersion of the Gaussian lines are those of M1 in Table~\ref{tbl:avrg_par}. Quoted errors are at the 90\% confidence level.
\end{list}

\end{table*}

\begin{figure} \centering{ 
\includegraphics[bb=0.6cm 1.2cm 10.7cm 28cm,clip=true,angle=-90,width=8.5cm]{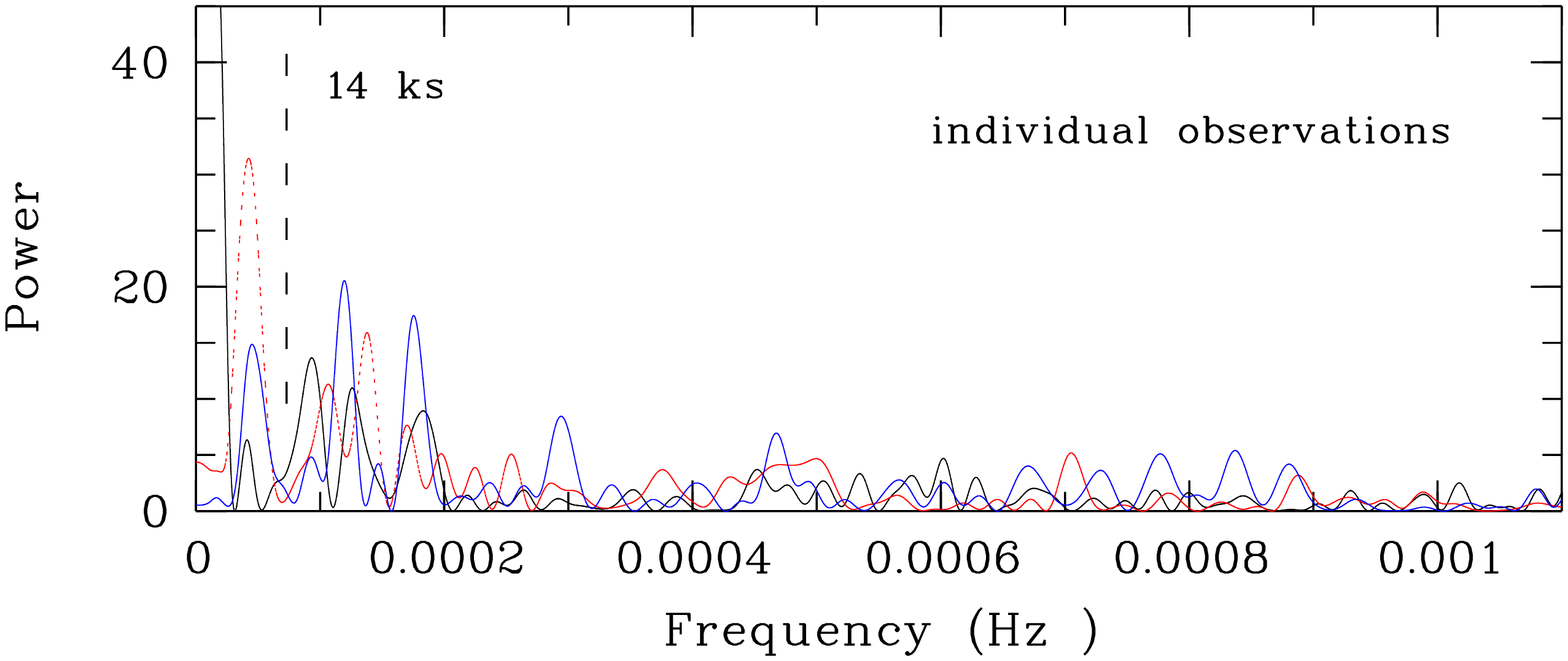}
\includegraphics[bb=1.3cm 1.2cm 13cm 28cm,clip=true,angle=-90,width=8.5cm]{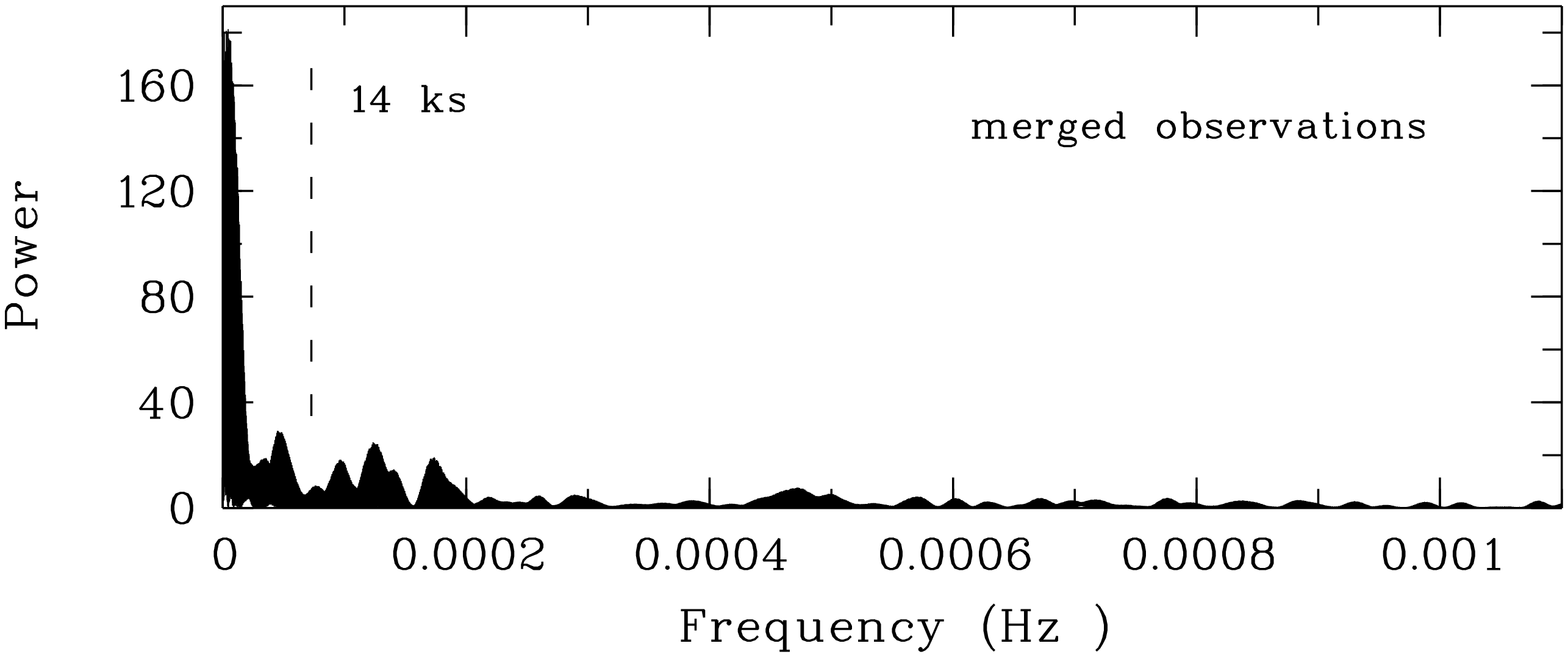}
\caption{Power spectra at low frequencies from 0.6--12 keV light curves binned to 150 s. Top: from individual observations (black: \obsA, red: \obsB, and blue: \obsC); bottom: from the merged 1, 2, and 3 observations. 
\label{fig:scargle}}  } \end{figure}

The Z$^{2}_{n}$ tool fails to detect any significant coherent signal in the frequency range 0.005--2.5\,Hz. The upper limit on the pulsed fraction is $\sim$ 2.5\% (at 0.6--12 keV). Although quasi-periodic variability exists during each epoch on time scales of 300\,s up to several ks, Figure \ref{fig:scargle} (top) shows that no peak is common to all \XMM\ observations.  
Figure \ref{fig:scargle} (bottom) also shows the {\it Scargle} power spectrum of the three observations merged in a single time series.
There is no strong coherent low frequency signal.
In particular, we do not find any power at the 14\,ks period found by \citet{TO01} in the BeppoSAX data. 

As visible on Figure \ref{fig:lc_hrd}, the hardness ratio exhibits long period oscillations with a pattern of variability apparently
uncorrelated with that of the total flux.
It is worth noting that the Scargle hardness ratio power spectra of \obsA\ and \obsC\ both show peak at 14 ks, not detected in \obsB. But the reality of this 14 ks peak is doubtful given the absence of a similar feature in the power spectra of soft and hard light curves.

\begin{figure}\centering{ 
\includegraphics[bb=0.5cm 5.5cm 20.6cm 25cm,clip=true,width=9cm]{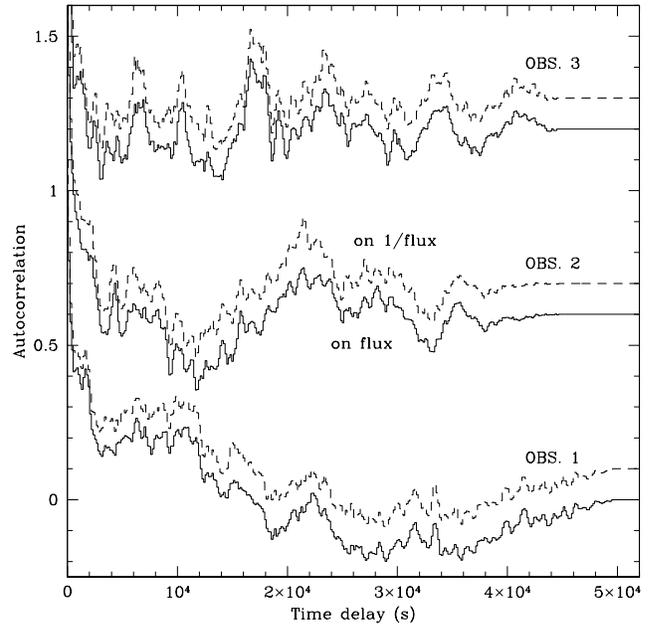}
\caption{Autocorrelation from 0.6-12 keV light curves (solid lines) and from their reciprocal fluxes, added to constant terms.} 
\label{fig:autocor}  } \end{figure}

To check for recurrences in the emission of \hd, we have also calculated the autocorrelation function from light curves and their inversed fluxes -- in order to emphasize the times  of low flux (see Fig. \ref{fig:autocor}), using the {\it Autocor}/Xronos v5.21. No pattern is evident.

\subsection{Power density spectrum}

\begin{figure}\centering{ 
\includegraphics[bb=0.5cm 5.5cm 20.6cm 25cm,clip=true,width=9cm]{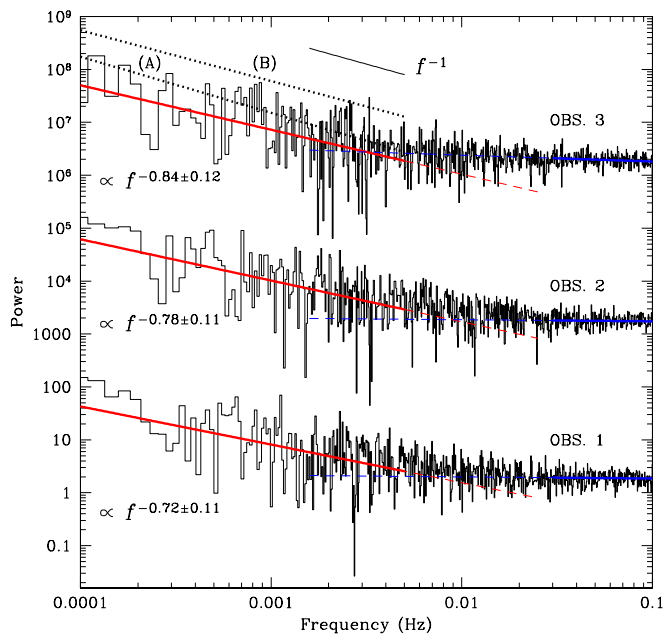}
\caption{Power spectra from 0.6-12 keV $pn$ data binned to 5\,s. For clarity, results
of the second and third observations were multiplied by factors of 10$^{3}$. A $f^{-1}$ profile is shown for comparison. The dotted lines (both offset by convenience) represent the $f^{-1.06\pm0.05}$ (A) and the $f^{-0.96\pm0.04}$ (B) profile derived from the \gcas's {\it RXTE} data on March 1996 and November 1998 \citep[after][]{RS00}, respectively.} 
\label{fig:pwsp}  } \end{figure}

We plot in Fig. \ref{fig:pwsp} the power spectra computed by {\it PowSpec}/Xronos v5.21 on the 0.6--12 keV $pn$ events, accumulated in time bins of 5 s. The power steadily rises at low frequencies, similar to what is observed in \gcas\ \citep{SRC98,RS00}. At frequencies below $\sim$ 0.003\,Hz, the spectra have the usual {\it power law} shape with an index of $\sim$ 0.72--0.84 up to a break at $f$ $\sim$ 0.01 Hz, where the white noise dominates. The {\it power law} index is significantly smaller than one -- and mutually consistent for all observations at $\pm$ 1$\sigma$ (see values in Fig. \ref{fig:pwsp}, computed from 10$^{-4}$ to 5$\times$10$^{-3}$ Hz). This indicates that the low frequency behaviour of \hd\ might be somewhat different from that of \gcas. The difference could lie either in the frequency of intermediate-timescale ($<$ few hours) flux ``undulations" or in the relative numbers of strong and weak flares.

\section{Discussion}

\subsection{\hd: a \gcas\ analog system}

\hd\ has X-ray and optical properties that are similar to those observed in 
\gcas. 
Contrary to ``normal" O-B stars which are usually soft \citep[$k$T $\sim$ 0.5 keV;][]{Berghofer96} and modest \citep[$L_{x}$ $\sim$ $10^{31-32}$\,\ergs;][]{Berghofer97} X-ray sources, \hd\ and \gcas\ emit preferentially hard X-rays at moderate luminosities.
In addition to the classical soft emission normally observed in B stars, 
a warm ($\sim$ 1--5 keV) component is needed to fit an otherwise dominant
hard continuum in \gcas\ 
\citep[$k$T $\sim$ 12 keV;][]{White82,Murakami86,Parmar93,Horag94,SRC98,Kubo98,Owens99,Smith04}.
Iron K lines (\FeHe\ and \FeH) as well as a fluorescent component 
at 6.4 keV are seen in emission. As for \hd, the light curve of \gcas\ displays strong variability on
time scales ranging from the photon limit of the instrument 
 to several minutes. These have the form of flare-like events \citep{Murakami86,SRC98,RS00}. 
The intrinsic luminosity of the {\it cool} component of \hd\ 
[$\sim$ (5--10)$\times$10$^{30}$ \ergs, in 0.1--2.4 keV] is 
comparable to the shocked-wind emission displayed by ``normal" B stars. 
On the other hand, the suspected variability by a factor of $\sim$ 5 in the unabsorbed soft-energy flux is in contrast with the X-ray
variability in this energy regime in ``normal" B stars.

The intensity and hardness variabilities, and the multi-temperature 
model needed to describe the X-ray emission of \hd, and \gcas\ 
\citep{Smith04}, strongly suggest a complex X-ray environment. 
We did not find any evidence for coherent ``pulsations'' in \hd.

We have noted that \hd\ and \gcas\ also display striking 
similarities in their optical spectra. The emission features in 
the yellow-red spectra confirm the information from
the Balmer line emissions that the Be disks are extensive and probably
dense. They also provide information on the thermodynamic description
and the viewing aspect of the disk. The appearance of migrating subfeatures
in the line profiles is similar to their appearance in the UV and optical
spectrum of \gcas. In the latter case one can also state that this is
consistent with the brief appearances of UV continuum absorptions that
seem to be formed by transiting
 clouds forced into corotation around the 
Be star \citep{SRH98} by a putative magnetic field and which are 
arguably anticorrelated with X-ray variations having several hour 
timescales \citep{SRC98}. 
As noted above, the 130 day modulation of the 2002
optical light may well be an analog of the now well-documented optical 
cycles of \gcas. These in turn correlate well with 3$\times$-amplitude 
X-ray cycles in \gcas\ \citep{RSH02,SB06}.

For both \hd, and \gcas, the ISM column density is much less than columns derived from modeling the soft X-ray spectra, and therefore the
most of the column density to the X-ray source is due to local absorptions.

Altogether, the set of X-ray and optical properties of \hd\ confirm its classification as a member of the newly established class of \gcas-like stars as proposed by \citet{Motch06a} and \citet{Lopes06}.

\subsection{Spectral changes}

Our spectral and timing analysis reveal the complex behaviour of the X-ray emission of \hd\ on short ($\sim$ tens to thousand of seconds) as well
as on long time scales ($\sim$ months--year). 

Strong and rapid variabilities are clearly observed in soft and hard light curves, even though apparently noncoherent, on time scales as short as $\sim$ 10 s. Such shots, exhibiting increases up to $\sim$ 100\% in flux, are superimposed over a long-term ($\sim$ 5--10 ks) trend of the X-ray emission,  but their occurence and intensity do not depend upon whether the source is in a ``low" or ``high" state. 
This chaotic behaviour strongly suggests the existence of numerous non-cospatial X-ray active regions, resulting in shots through flare-like events. In this picture, the basal component would be due to an averaged emission of a collection of a large number of such active regions.
Actually, the best fitting 3-T models show that several distinct  emitting regions of different temperatures, emission measures and fluxes, and probably column densities contribute to the overall X-ray energy distribution on large time scales. The variability of these parameters in different putative sites seems so far uncorrelated. 

According to M1, the X-ray emission of \hd\ is dominated by a {\it hot}-thermal component, which accounts for $\sim$ 80\% of the unabsorbed 0.2--12 keV flux. Its {\it warm} and {\it cool} components represent the other $\sim$ 18\% and $\sim$ 2\%.
The unabsorbed X-ray flux of \hd\ has steadily decreased over the two first \XMM\ observations, increasing during the last observation. This trend was apparently accompanied by a slight hardening of the X-ray spectrum. 

The EM of the {\it hot} component seems to decrease by $\sim$ 10\% over the subsequent observations, while the EM of the {\it cool} and {\it warm} components follow a more complicated pattern, which is anti-correlated one with another. If the different plasmas have similar densities, our results show that the total emitting volume in all observations is dominated by the {\it hot} component ($\sim$ 70--80\%), with a small contribution of the {\it warm} ($\sim$ 15--25\%) and {\it cool} (few percent) components.

The EPIC energy band (0.2--12 keV) does not extend enough in the hard X-rays to really constrain the high temperatures ($\ga$ 20 keV) derived for the hot emission of \hd. The fits yield large errors, and the temperatures in each epoch are the same within the 90\% confidence limit. However, the varying relative intensity of the ionized iron lines (Fig. \ref{fig:fekalpha}) suggest that the physical parameters of the X-ray environment has changed in the different epochs.

\subsection{Origin of the X-ray emission}

In spite of several X-ray campaigns on \gcas, the true nature of its X-ray emission remains an open question: accretion onto a degenerate companion or magnetic active Be star?
This controversy itself extends to \hd\ and all \gcas-analog systems.

\subsubsection{Be/X-ray binary?}

Accretion onto a degenerate companion star has been invoked 
to explain the anomalous, variable X-ray flux of \gcas, 
mainly due its hard emission and moderate luminosity -- of about 
10$^{32-33}$\,erg\,s$^{-1}$ in 0.2--12 keV. This could be easily obtained 
if matter is falling directly onto a NS either from the Be star's wind
or its decretion disk
(\Macc\ $\sim$ 3$\times$10$^{-13}$ 
\Msol\,yr$^{-1}$), or conceivably (though not easily)
a WD (\Macc\ $\sim$ 3$\times$10$^{-10}$ \Msol\,yr$^{-1}$). 
Following the hard-thermal nature of its X-rays, 
typical of CVs and in contrast with the non-thermal emission detected in 
all well-known classic Be + NS systems, and despite the lower yield in luminosity, a WD is preferred instead of a NS. 

Be + WD systems are predicted by several massive binary
evolution models to be the outcomes 
describing the evolution of massive binary systems.
This configuration is expected for all B types, with a distribution peaking around the B2--3 types \citep{Pols91}.
Typical models predict that these systems should comprise about 
70\% of the total evolved Be binary outcomes
\citep{vandenHeuvel87,Waters89a,Pols91,VanBever97,Raguzova01}. However, we
note that this fraction depends sensitively on the final mass of the 
B primary such that it is low for evolved systems with early B primaries.
Thus, in these computations all known Be/X-ray binaries 
have a companion NS \citep[see][for a recent compilation of Be/X-ray 
properties]{Raguzova05}. So far no 
Be + WD system has been identified. 
On the basis of their X-ray properties, notably their (i) hard-thermal 
spectra, (ii) iron lines, and (iii) luminosities, 
that are reminiscent of similar properties in some CVs, the \gcas\ analogs 
are tenable Be + WD candidates.
However, one difference we can point out is that
the absence of a continuous differential emission measure and the variable energy distribution of \hd\ discourage the expectation of a cooling flow model often associated with the boundary layer in CVs and therefore it is an argument against the presence of an accreting white dwarf. 
However, this argument may be made on a dearth of examples of high
quality spectra for which a DEM can be absolutely ruled out instead of
a multi-temperature component model. [But this by itself does not necessarily rule out a
WD-accretion model.]
Another possible example of a Be + WD system might be the super-soft X-ray source XMMU J052016.0--692505, 
associated to a B0-3e star in the LMC, as proposed by \citet{Kahabka06}. 
Its blackbody emission, with bolometric luminosity 
$\ga$~10$^{34}$ erg\,s$^{-1}$, 
was interpreted as being due to a WD, 
accreting from a decretion disk of a Be star in a putative binary system.

In CV systems a WD, strongly magnetized or not, accretes from a Roche lobe-filling, late-type evolved companion star. An accretion disk is commonly formed in non-magnetic systems. However, it tends to be inhibited by the magnetic field lines in strongly magnetized ($\sim$ 10$^{10}$ G) stars. In both types of systems accretion shocks onto the WD or the boundary layer between the WD's surface and a putative disk will result in a X-ray spectrum characterized by plasma temperatures ranging from a few keV to tens of keV. Their luminosities are moderately high ($\sim$ 10$^{32-33}$ \ergs). Interestingly, their X-rays are typically modeled as emissions of plasmas having at least a few components and indeed perhaps a broad continuous temperature distribution. The latter possibility is typical of a cooling process, e.g. in which the emission measure of each temperature component scales with the temperature as $d$EM/$d$T $\propto$ T$^{\alpha}$. 

The prominent ionized iron lines, \FeHe\ and \FeH, and the fluorescent 6.4 
keV iron feature, present in \hd\ and in all \gcas\ analogs so far are also 
reminiscent of those detected in several CV systems, with similar EWs and 
intensities \citep{Hellier98,Ezuka99}.
Curiously, \citet{Ezuka99} noted that the ionization temperature 
derived from Fe\,K lines in CVs are systematically lower than that derived 
from continuum, as we found for \hd\ (Section \ref{sec:feka_cont}).

In fact, \gcas, the prototype of this class, itself is a component of
 a binary system \citep{Harmanec00,Miroshnichenko02}. The companion star has
roughly 1\,$M_{\odot}$, but its nature, degenerate or otherwise, is unknown.
In the case of of \gcas\ itself, its high mass of about
15\,M$_{\odot}$ \citep[e.g.][]{Stee95,Harmanec00}
places constraints on binary evolution scenarios ending
up with a massive Be star with a WD companion. However, these
constraints are less important if the mass of HD\,110432 is
only 9.6\,M$_{\odot}$ \citep{Zorec05}.

The binary status of \hd\ and all other \gcas-analog candidates is unknown.
If \hd\ actually belongs to the $\sim$ 60 Myr old open
cluster \object{NGC\,4609} \citep{Feinstein71}, then the
largest age for a 9.6\Msol B1IVe star of $\sim$ 20 Myr would suggest that \hd\ is a blue straggler.
Interestingly, two other \gcas\ candidates, the star in
\object{NGC\,6649} \citep{Marco06} and \hdb\ \citep{SH07}
also appear to be blue stragglers. Therefore, an evolved
status may be a prerequisite to the source being a peculiar
X-ray emitter. An advanced age could hint at a few scenarios
for the X-ray production, such as an accreting binary
companion, or a Be star with a strong surface field
resulting from the buoyant rise of flux through the star's
radiative interior \citep[e.g.][]{MacGregor03}.

It is also not yet clear
that a WD in a Be binary system as widely spaced as \gcas\
could accrete enough material from the wind to account for the observed X-ray luminosity.
In particular, the tidal torquing of the Be disk by the secondary
star may effectively truncate the disk inside the secondary's orbital
radius \citep{Okazaki01}, 
which would result in too little mass accretion to be important. 
Interestingly, there is no detected correlation between X-ray flux and 
the orbital phase of \gcas\ \citep[RXTE observations;][]{RSH02}, or outburst. 
This could be explained if the accreting star were 
in a nearly circular orbit close to the plane of the Be decretion disk.

For completeness, we point out that
the \gcas-like systems \hda\ and \hdb\ exhibit 3.2 ks 
\citep[\XMM;][]{Lopes06} and 1.5 ks \citep[\XMM\ and Chandra;][]{SH07} 
oscillations, respectively. 
If these should turn out to be stable, they could be the signature 
of a spinning compact object.

\subsubsection{Magnetic active Be star?}

Evidence is accumulating in favor of magnetic activities 
in \gcas\ and \hd, that could produce the X-ray emission in these 
objects. The fundamental question of whether these stars are 
magnetic with a complex but stable field topology seems less hypothetical with the discovery of a coherent
variation over 9 years with a photometric period of 1.21 days 
in the optical $B, V$ passbands \citep{SB06}. This periodicity
appears to be best explained by a feature on the star's 
surface, since 1.2 days is consistent with the star's expected radius, rotational velocity, and 
obliquity. Perhaps also 
relevant are the recurrent blue-to-red 
``migrating sub-structures" ({\it msf}) seen in the line optical 
profiles of both stars often \citep{Yang88, 
SR99,SB06}. These features are reminiscent of the active and
rapidly rotating pre-main sequence K star AB Dor \citep{Collier89}. 
The {\it msf} are most easily 
interpreted as being due to corotating clouds 
anchored onto the stellar surface by magnetic confinement \citep{SRH98}.
In addition, the $\sim$ 3--4\% sinusoidal
modulations in the $B$ and $V$ photometric bands 
with a timescale of 130 days detected in \hd\ \citep{SB06}
are reminiscent of the optical cycles reported in \gcas, which are
best explained by variations in the integrated flux of the Be disk
\citep{SRH98,SHV06}. 

The key assumption of the scenario in which X-rays are due to 
magnetic activity is that the Be star has a surface magnetic 
field some fraction of which threads into 
the inner regions of its ionized circumstellar
disk \citep[][and references therein]{RSH02}. According to
the hypothesis of \citet{RSH02}, field entrainment 
produces turbulence within the disk and the difference in angular 
velocities between the disk and the star stresses and shears 
magnetic lines. Magnetic reconnection leads to the ejection of 
high energy particles which impact the Be star and generate hard 
X-ray emission.  This scenario is supported by UV observation
of highly redshifted line absorptions in \hd\ that could be manifestations
of accelerated material \citep{SR99}. 
In this context, $\gamma$\,Cas analogs may be related to magnetic Op 
and Bp stars like $\theta^1$\,Ori\,C \citep{Donati02}. 
[However, we note that the hottest X-ray plasma of $\theta^1$\,Ori\,C has a temperature $k$T of only $\sim$ 2.5 keV \citep{Gagne05} or $\sim$ 4.5 keV \citep{Schulz00}].
As part of a star-disc interaction, X-rays could 
be modulated by a dynamo-like process within the Be's 
circumstellar disk \citep{RSH02}. A dense circumstellar disk in \gcas\ 
analogs seems to be a necessary ingredient for the magnetic picture.

It is not yet clear how observations might
eventually provide quantitative constraints on the star-disc magnetic 
picture.  Similarly, the theory is not yet developed enough to predict
a characteristic temperature or to explain the presence of plasma
with more than one temperature.
According to the X-ray variabilities 
detected in \gcas\ analogs, the surface of these stars would
have to have a complicated magnetic topology.
In any case, the 1/$f$ tendency roughly observed in the periodogram at low frequencies is in agreement with stochastically appearing blobs. 
Indeed, at this point dynamo
models also cannot make sensible predictions for {\it any} astrophysical
environment (including the solar interior).

\section{Conclusions} \label{sct:conclusions} 

\hd\ is a member of the recently discovered class of \gcas-analogs 
\citep{Motch06a,Lopes06}, according to its X-ray and optical properties, summarized as follows. 
This class is composed of Be stars having X-ray properties described by moderately luminous and variable light curves ($\sim$ 10$^{32-33}$\,erg\,s$^{-1}$; 0.2--12 keV) and spectra indicative of multiple thermal components dominated by a hard component \citep[$k$T\,$\ga$\,7 keV; current limit determined by \hda;][]{Lopes06}.
No X-ray outbursts have been observed.
The thermal origin of their X-ray emissions is strongly supported by the presence of an \FeK\ complex.
The presence of a fluorescent iron line at 6,4 keV suggests a dense-cold medium close and in the line-of-sight of the X-ray source, or fluorescence over the surface of a accreting WD companion.
Curiously, all identified members are B0.5-B1 stars with very large and 
nearly symmetrical H$\alpha$ profile, and \ion{Fe}{ii} emission lines, supporting 
dense and/or large and likely stable circumstellar disks. 
{\it In toto,} these features are not exhibited in optical and X-ray spectra and light curves of ``normal"
Be stars or known Be/X-ray binaries.

The X-ray spectrum of \hd\ is very complex and variable. 
It probably results from 
a composition of three thermal plasmas ($k$T $\sim$ 0.2--0.7 + 3--6 + 16--37 keV), and thus \hd's X-ray temperature is the hottest known for any Be star. 
A mixed model of thermal and non-thermal components, resulting in a dominant plasma of $k$T $\sim$ 8--11 keV and a power law description for the high energy tail in \hd\ is statistically acceptable. However, it appears a less realistic interpretation because of the presence of the \FeH\ Ly$\beta$ line.
Future X-ray observations carried out by Suzaku observatory, covering simultaneously from 0.2 to 70 keV with the moderate resolution available by the XIS and HXD cameras, would be useful to put additional constraints on the nature of the hard energy tail in \hd. 
We would emphasize that the increase of the sample size makes it more
likely that one of these stars will be able to undergo an optical Be outburst
cycle. A monitoring of X-ray characteristics during these cycles would greatly
clarify the role of the Be disk in the production of these unique (for
Be stars) hard X-rays.

In addition to the iron lines detected in \gcas-analogs, a variable \FeH\ 
Ly$\beta$ emission line at 8.2 keV is detected for \hd, in agreement with 
its hot and variable X-ray component. Notably, its light curve exhibits 
rapid and strong variations, in soft and hard X-ray bands throughout 
the 0.2--12 keV range. These variations are followed by long term variations 
and (as for \gcas) ubiquitous flare-like events in the source's hardness which recur on timescales as short as detectable with current X-ray instrumentation. There is no evidence for high-frequency pulsations, but our 
low-frequency analysis reveals unstable long-term variations on timescales of a few hours.

Our X-ray spectral analysis is consistent with the presence of
an accreting white dwarf for \hd, on the basis of certain spectral similarities with CVs. Similarly, the X-ray and optical characteristics of \hd\ are consistent with the magnetic star-disk interaction scenario, proposed for \gcas\ by 
\citet{RSH02} and references. The presence or absence of a companion star to
\hd\ and if appropriate the description of its orbit, or additional indications of magnetic fields will provide the 
means to make substantial progress in elucidating the processes behind 
its X-ray emission.

\begin{acknowledgements}

We are grateful to the anonymous referee for his/her valuable comments. R.L.O.  acknowledges financial support from Brazilian agencies FAPESP (grant 03/06861-6) and CAPES (grant
BEX0784/04-4), and the Observatoire de Strasbourg (CNRS).
I.N. is a researcher of the 
 programme {\em Ram\'on y Cajal}, funded by the Spanish Ministerio de 
 Ciencia y Tecnolog\'{\i}a (currently Ministerio de Educaci\'on y 
 Ciencia) and the University of Alicante, with partial 
 support from the Generalitat Valenciana and the European Regional 
 Development Fund (ERDF/FEDER). 
 This research is partially supported by the MCyT (currently MEC) under 
 grant AYA2005-00095.

\end{acknowledgements}

\end{document}